\documentclass{PoS}
\usepackage{subfigure}

\title{Lattice QCD: a critical status report}

\ShortTitle{Lattice QCD: a critical status report}

\author{
\begin{flushright}
DESY 08-138
\end{flushright}
\speaker{Karl Jansen}\\
        DESY, \\
        Platanenallee 6, 
        15738 Zeuthen\\
        E-mail: \email{Karl.Jansen@desy.de}}

\abstract{The substantial progress that has been achieved 
          in lattice QCD in the last years is pointed out. 
          I compare the simulation cost and 
          systematic effects of several lattice QCD formulations and discuss   
          a number of topics such as lattice spacing scaling, applications
          of chiral perturbation theory, non-perturbative renormalization and 
          finite volume effects. Additionally, the importance of 
          demonstrating universality is emphasized.} 

\FullConference{The XXVI International Symposium on Lattice Field Theory\\
		 July 14-19 2008\\
		 Williamsburg, Virginia, USA}

\begin{document}

\section{Introduction}

At the Capri lattice symposium in 1989, 
it was stated that in lattice field theory it would have been necessary 
{\em ``both a $10^8$ increase in computing power AND spectacular 
algorithmic advances before a useful interaction with experiments starts 
taking place.''}
\cite{Wilson:1989ax}.
At the time of this statement, in 1989, the available computing power was around 
$10-100$Gigaflops \cite{Tripiccione:1989jy}. As a consequence, lattice field theory 
would have needed at least Exaflops computers in order to perform realistic 
simulations and to produce any experimentally 
interesting output. 

In addition, at the lattice conference in Berlin in 2001 a serious 
attempt to determine the scaling behaviour of the 
algorithms to simulate lattice QCD as a function of the quark mass, 
the lattice spacing and the volume was  
made. It was found 
\cite{Bernard:2002pd,Jansen:2003nt}
that the expense of lattice QCD simulations increases with a large inverse power 
of the quark mass leading to exorbitant costs at the physical value of the 
pseudo scalar mass, which we will denote as the physical point further on. 
In fact, the simulations costs turned out to be  
already very large much before being able to reach the physical point 
such that  
simulations with pseudo scalar masses below, say, 300MeV 
seemed to be completely out of reach. 

However, in stark contrast to the above rather pessimistic scenario, 
it could be witnessed 
at Lattice 2008 in Williamsburg that a number 
of lattice QCD simulations with pseudo scalar masses well below 300MeV, values 
of the lattice spacings down to $a\approx 0.05$fm and box sizes with linear 
extent $\gtrsim 2.5$fm are currently being performed by various 
international collaborations.
Such simulations allow then for an extrapolation of the results to the 
physical point and to the continuum limit while keeping also the 
finite volume effects
under control. 
And, there are even more ambitious simulations starting presently which 
are performed at or very close to the physical point \cite{pacscs2008}. 


\begin{figure}[t]
{\includegraphics[width=0.95\linewidth]{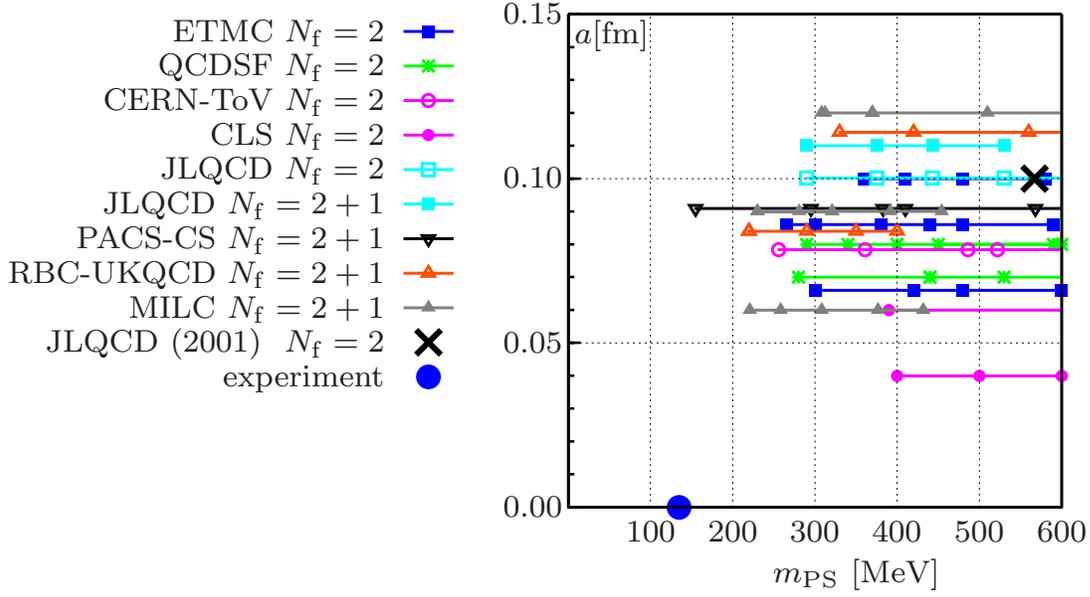}}
\caption{The values of the lattice spacing $a$ and pseudo scalar masses 
$m_\mathrm{PS}$ as
employed presently in typical QCD simulations by various collaborations as 
(incompletely) listed in the legend. The blue dot indicates the physical point
where in the continuum the pseudo scalar assume assumes its experimentally measured
value. The black cross represents a state of the art simulation by the JLQCD 
collaboration in 2001.}
\label{fig:parameters}
\end{figure}

Thus, the prognosis which emerged in 1989 has {\em not} been fulfilled:
already nowadays completely realistic simulations of lattice 
QCD are possible on available machines delivering a few 100 Teraflops. 
The values of the lattice spacings and pseudo scalar masses which are 
employed in todays simulations are compiled in fig.~\ref{fig:parameters}.
In the figure, the blue dot indicates the physical point. The black cross
represents a state of the art simulation in the year 2001. As can be seen 
in the graph, most of the simulations now go well beyond what could be 
reached in 2001 demonstrating clearly the progress in performing realistic 
simulations.

This phase transition-like change in the situation is due to three 
main developments: 
$i)$ algorithmic breakthroughs that either shifted the wall of the 
algorithm scaling in the quark mass or even changed this scaling behaviour  
itself 
drastically, $ii)$ machine 
development; the computing power of the present BG/P systems is even outperforming 
Moore's law, $iii)$ conceptual developments, such as the use of 
improved actions which reduce lattice artefacts
and the development of non-perturbative renormalization. 

To illustrate the status of present lattice QCD simulations
let me give just two examples for the results obtained at the moment. 

\subsection{Baryon spectrum} 

When simulations of lattice QCD were started, the computation of the baryon spectrum 
was one of the main goals. 
Although such a computation can only be considered 
as a post diction since the masses are measured precisely in experiment, their 
determination on the lattice has always been considered as an 
important  benchmark calculation. 

\begin{figure}[t]
{\includegraphics[width=0.80\linewidth]{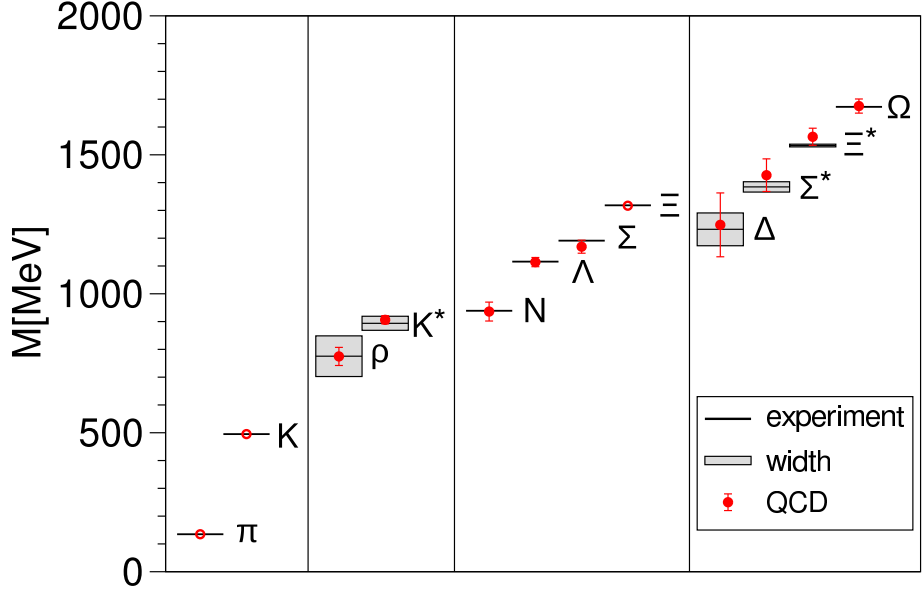}}
  \caption{The Baryon spectrum as obtained by the Budapest-Marseille-Wuppertal
   collaboration \cite{bmw2008}.}
  \label{fig:baryonspectrum}
\end{figure}

It is very reassuring that many international collaborations working with 
lattice field theoretical methods are either very close to finish the computation 
of the baryon spectrum 
\cite{Alexandrou:2008tn,Aubin:2004wf,Bernard:2007ux,Allton:2008pn,Aoki:2008sm,WalkerLoud:2008bp,Gockeler:2007rx}
or, as in the case of the Budapest-Marseille-Wuppertal 
collaboration have already accomplished the goal \cite{bmw2008}. 
In fig.~\ref{fig:baryonspectrum} 
the recent results from the BMW-collaboration presented at this 
conference is shown. 

In order to obtain the baryon spectrum shown in the graph, simulations at 
three different
values of the lattice spacing $0.065\mathrm{fm} \lesssim a \lesssim 0.125\mathrm{fm}$ 
have been performed. The values of the pseudo scalar 
masses are bracket by $200\mathrm{MeV} \lesssim m_\mathrm{PS} \lesssim 500\mathrm{MeV}$. 
Finally, the box size has been chosen such that
$m_\mathrm{PS}L\gtrsim 4$. 
This setup allows for extrapolations to the physical point.  
It also allows for a continuum limit extrapolation and suppresses finite volume 
effects for many quantities. 
Thus, the spectrum calculation shown in fig.~\ref{fig:baryonspectrum} can  
be considered as a first 
lattice benchmark calculation with, however, 
the caveat of the need for an eventual 
cross-check.  
Nevertheless, the agreement of the lattice results with the experimentally measured 
Baryon spectrum is highly non-trivial.                             

The work of ref.~\cite{bmw2008} is a lattice computation 
from only one group and from 
only one lattice discretization. In order to say with confidence that 
this is a direct non-perturbative {\em QCD} result, it is mandatory,  
in my opinion,
that the computation is repeated by at least one different 
collaboration with 
most preferably a different lattice action. 
Only then we will have demonstrated that lattice methods 
provide a reliable tool to 
obtain physical results from first principles and in a non-perturbative 
fashion.

The reason for additional calculations of physical quantities
is that different lattice formulations of QCD will show different
systematic errors and only the continuum limit will reveal whether consistent 
results are obtained, thus demonstrating non-perturbatively that  
universality is realized. This point is further 
discussed below. There it will be demonstrated that for the baryon masses 
different discretizations indeed seem to give the same continuum limit values. 
However, for other quantities the situation is much less clear which presumably 
just means that we need to understand better the inherent systematic effects in 
our lattice simulations. 

\subsection{Low energy constants}

Another field where a substantial progress could be achieved is the 
determination of  
low energy constants of chiral perturbation theory. In the past such determinations
were blocked by the expense of performing 
dynamical fermion simulations with pseudo scalar 
masses of 300MeV or lower. 

With the advances in lattice field theory in recent years, 
pseudo scalar mass values of $m_\mathrm{PS}\approx 300$MeV are  
simulated today by a number of collaborations 
as shown in fig.~\ref{fig:parameters}. 
In particular, many collaborations
now have very precise results for the pseudo scalar masses and decay constants for
$250\mathrm{MeV}\lesssim m_\mathrm{PS} \lesssim 450\mathrm{MeV}$.                     
The existing data show strong indications, 
at least for the case of $N_f=2$ flavours of quarks, that chiral perturbation 
theory is applicable in this regime of corresponding quark masses. 

\begin{figure}[t]
  \subfigure[Results from the European twisted mass collaboration (ETMC) 
             comparing NLO and NNLO chiral perturbation theory fits 
            to their numerical data at two values of the lattice 
            spacing, $a\approx 0.085$ fm ($\beta=3.9$) and $a\approx 0.075$ fm 
            ($\beta=4.05$). In the fit region, covering pseudo scalar 
            masses between 250MeV and 450MeV no sensitivity to the NNLO 
            correction can be detected.
  \label{fig:lec1}]
  {\includegraphics[width=0.50\linewidth]{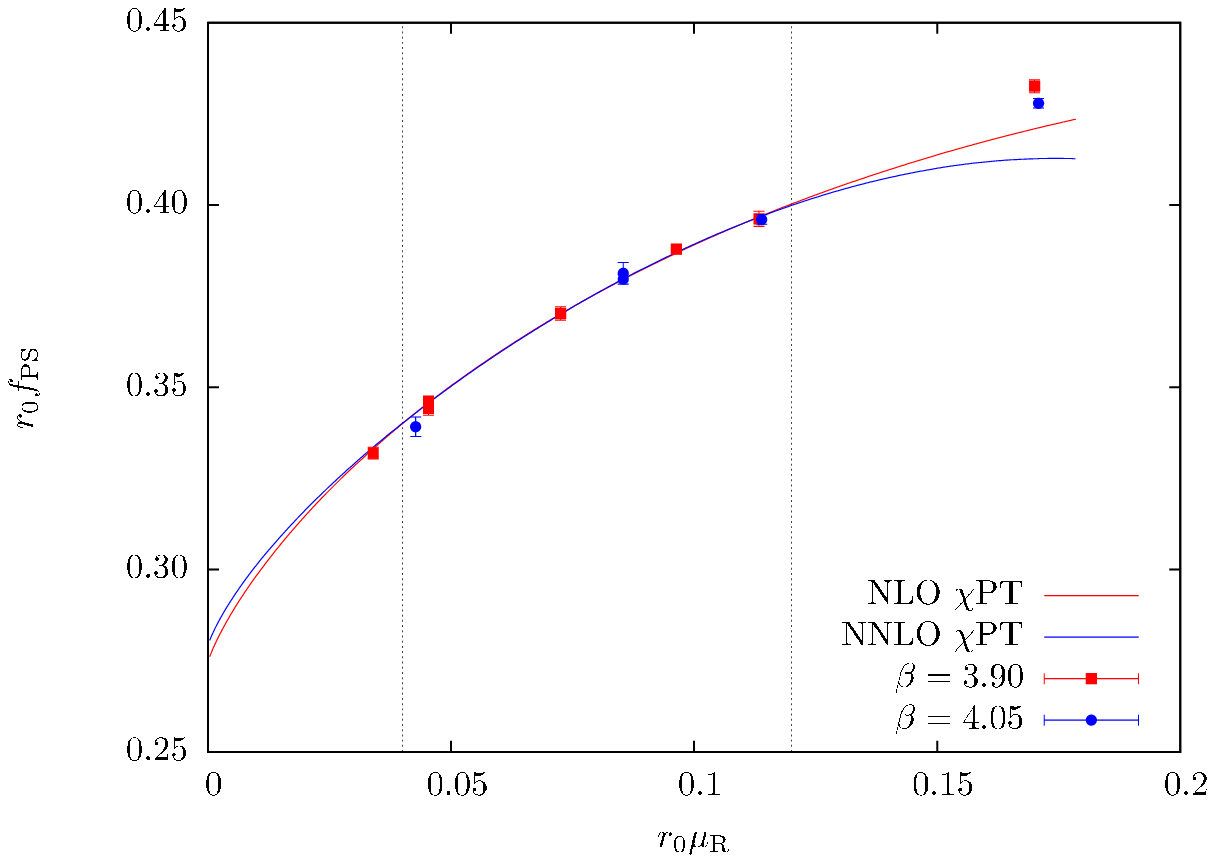}}\quad
  \subfigure[Results from the Japanese lattice QCD (JLQCD) collaboration. 
   The different expansion parameters are: 
   $x=\frac{2B_0m_q}{(4\pi f)^2}$, $\hat{x}=(\frac{m_\pi}{4\pi f})^2$ and
   $\xi=(\frac{m_\pi}{4\pi f_\pi})^2$. \label{fig:lec2}]
  {\includegraphics[width=0.46\linewidth]{fps_nlo.eps}}
  \caption{Confronting lattice QCD results for the pseudo scalar 
  decay constant with chiral perturbation theory.}
  \label{fig:lec}
\end{figure}

Thus, fits to formulae from chiral perturbation to the very accurate numerical data
allow for the determination of the low energy constants of chiral 
perturbation theory with a high precision. 
In fig.~\ref{fig:lec} two examples for fits to 
formulae from chiral perturbation theory are given. The first example 
is from the
European Twisted Mass collaboration (ETMC) 
\cite{Boucaud:2007uk,Urbach:2007rt,Dimopoulos:2008sy}.       
It shows the pseudo scalar decay constant as a function of the renormalized 
quark mass, both in units of $r_0$. In the range of the fit, indicated
by the two vertical dotted lines, both, the next to leading order (NLO) 
and the next to next leading order (NNLO) curves are
shown. There is no sensitivity to the NNLO corrections and the NLO
formula describes the data very well. 

In fig.~\ref{fig:lec2} another example, taken from the Japanese lattice QCD (JLQCD) 
collaboration \cite{Noaki:2008gx}, is given again for the case of the 
pseudo scalar decay constant.
A comparison is made using different expansion parameters for the chiral fit 
formula. For pseudo scalar masses of $m_\mathrm{PS} \lesssim 500$MeV all fits 
agree indicating again that for such a range of pseudo scalar masses chiral 
perturbation theory is applicable. We will discuss chiral perturbation theory 
fits and possible problems below again.

\section{Cost of simulation}

\begin{figure}[t]
\subfigure[A comparison of the cost estimate taken 
from ref.~\cite{Urbach:2005ji}. The solid line \cite{Orth:2005kq} 
(indicated as ref.~12 in the plot) 
indicates the cost of simulations around the time of 
the Berlin lattice symposium in 2001. 
The data represented by the filled squares are 
extrapolated with
$(m_\mathrm{PS}/m_\mathrm{V})^{-4}$ (dashed) and with
$(m_\mathrm{PS}/m_\mathrm{V})^{-6}$ (dotted), respectively. The arrow
indicates the physical pion to rho meson mass ratio. Additionally,
points from staggered simulations were used for the
corresponding plot taken from ref.~\cite{Jansen:2003nt}.
\label{fig:cost1}]
{\includegraphics[width=0.43\linewidth]{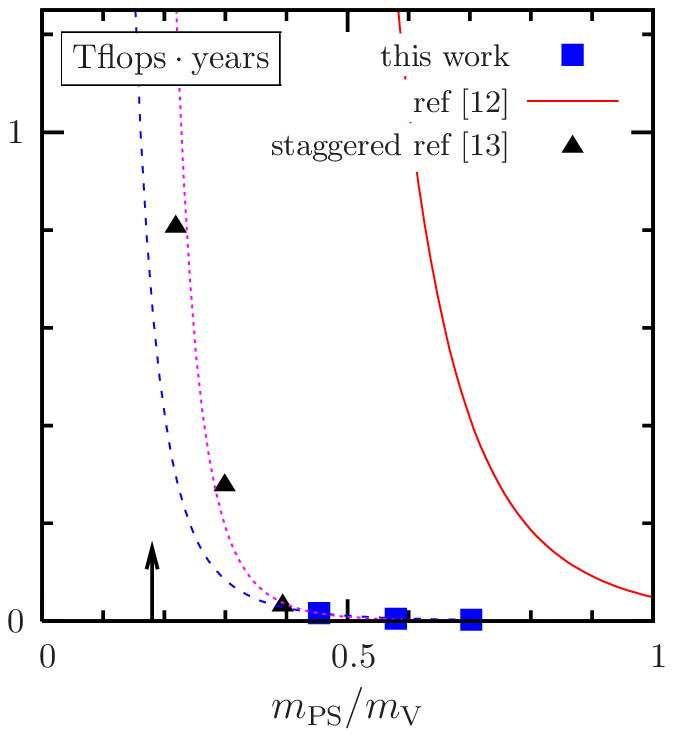}}\quad
  \subfigure[The cost of dynamical fermion simulations using different kind
of algorithms and lattice fermions. TM stands for twisted mass and data
are taken from \cite{Urbach:2005ji}. DW are domain wall fermions and the
performance figures are from \cite{christ:priv}. The Wilson performance
line is taken from ref.~\cite{DelDebbio:2006cn}, the Wilson performance
line using also the deflation technique of ref.~\cite{Luscher:2007es} is
shown as the dotted line. Finally, the staggered performance cost
\cite{toussaint:priv} using
the algorithm of ref.~\cite{Clark:2006wp} is represented by the lowest 
lying (blue) line.
\label{fig:cost2}]
{\includegraphics[width=0.50\linewidth]{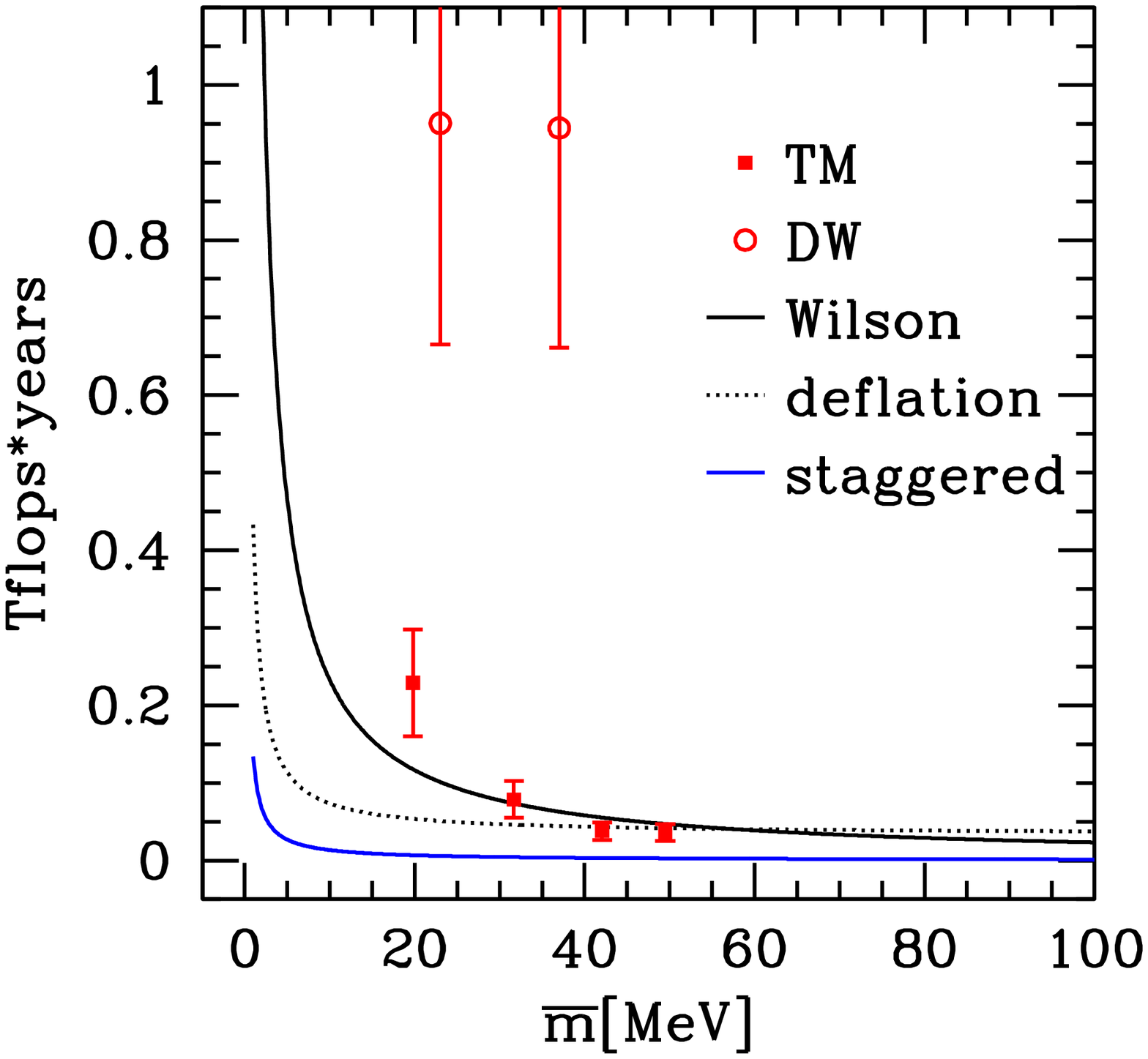}}\quad
\caption{The Berlin wall plots.}
\label{fig:cost}
\end{figure}

For sure, conceptual developments -- such as O$(a)$-improvement or non-perturbative 
re\-nor\-ma\-li\-za\-ti\-on -- and new supercomputer architectures 
are playing an important role
for the breakthrough advances in lattice QCD described above.
However, the major factor in this development 
is due to substantial advances in the algorithms that are used to 
perform our lattice QCD simulations.
In fig.~\ref{fig:cost1} we show the cost to produce 1000 independent 
configurations on a lattice of linear size of $L=2.1$fm  
with a value of the lattice spacing of $a=0.08$fm. 
Although the physical size of the considered box is, by today's standards, 
not very ambitious, it 
is chosen in order to compare with the situation at the Lattice symposium 
2003 in 
Tsukuba \cite{Jansen:2003nt}. There, it was shown that a  
Wilson fermion simulation at a renormalized quark mass 
of about 20MeV (in the $\overline{\mathrm{MS}}$-scheme at scale 
2GeV) would have needed
an unrealistic amount of computer resources. 
The progress that took place in the last years 
is illustrated  
in fig.~\ref{fig:cost1}. 
Note that all the cost data were scaled to match a lattice
time extend of $T/a=40$. 
In fig.~\ref{fig:cost1} it is also shown that simulations
with staggered fermions were much faster in 2003 than corresponding Wilson fermion
simulations. 

The situation as of today is summarized in fig.~\ref{fig:cost2}.
The red squares in the graph of fig.~\ref{fig:cost1}  
correspond to measured performance costs 
from maximally twisted mass fermions (TM) using the algorithm 
described in \cite{Urbach:2005ji}. These costs compare nicely with 
the performance figure for Wilson fermions using the DD-HMC 
algorithm \cite{Luscher:2005rx} shown as the solid 
black line which uses the cost formula, 

\begin{equation}
  C_\mathrm{op} = k\ \left( \frac{20\ \mathrm{MeV}}{\bar{m}}\right)^{c_m}\
  \left(\frac{L}{3\ \mathrm{fm}}\right)^{c_L}\ \left(\frac{0.1\
      \mathrm{fm}}{a}\right)^{c_a}\
  \mathrm{Teraflops}\times\mathrm{years}
\label{costformula}
\end{equation}
with parameters as given in ref.~\cite{DelDebbio:2006cn}.
In eq.~(\ref{costformula}), $\bar{m}$ is the renormalized quark mass 
at a scale of 2GeV in the $\overline{\mathrm{MS}}$-scheme.
Typical values for the exponents in this formula are 
$c_m=1-2$, $c_L=4-5$ and $c_a=4-6$. Note that these values have 
a large uncertainty and should here only be taken as a guideline. 
The prefactor $k$ is typically $O(1)$ for Wilson fermions using 
the algorithms described in \cite{Luscher:2005rx,Urbach:2005ji} and 
$O(0.01)$ for staggered fermions \cite{toussaint:priv} when the algorithm 
of ref.~\cite{Clark:2006wp} is employed. 
The performance results for Wilson 
fermions using the above mentioned algorithms 
show a tremendous gain when compared to the situation in  
2003 \cite{Jansen:2003nt}, see fig.~\ref{fig:cost1}.

%

However, this is not even the end of the story. The dotted black line in
fig.~\ref{fig:cost2} shows the effect of using in-exact eigenvalue 
deflation of the lattice Dirac operator as described 
in ref.~\cite{Luscher:2007es}. As can be observed, the cost is almost flat 
as a function of 
the quark mass and the wall-like behaviour sets in only at values of the quark 
mass below 5MeV. This striking result is even beaten by simulation costs of
staggered fermions \cite{toussaint:priv} 
which are again a noticeable factor below the cost of the best 
Wilson fermion simulation. It should be stressed that the lines  
representing deflated Wilson and staggered fermions are fitting curves 
that are based on measured performance costs for values 
of $\bar{m}\gtrsim 20$MeV only. For completeness, in the graph the 
simulation costs \cite{Christ:2006zz,Allton:2008pn}
of domain wall fermions are also plotted \cite{christ:priv}. 
As can be seen, this formulation
of lattice fermions, although requiring an extra dimension, is only moderately 
more expensive than the one for Wilson formulations. 
Note that in principle deflation techniques can also be applied to 
twisted mass, domain wall 
and staggered fermions, leading possibly to similarly large 
gains as for Wilson fermions.

In conclusion, the Berlin Wall that was frightening the lattice community 
in 2001/2003 has been shifted to such small values of the quark mass
that for all practical simulations a realistic amount of computer time 
is needed which matches the capacity of modern supercomputers such 
as BG/P. (See \cite{Ishikawa2008} for an overview of present 
supercomputer architectures.) 
Typical physical situations of today are 
boxes with $L=3$fm and pseudo scalar masses of $200$MeV or 
even $140$MeV. Living in a time where a number of machines 
are available that reach several hundreds of Teraflops or even Petaflops, we will
see therefore in the near future many precise and phenomenologically
relevant results from the lattice.
Of course, if physical problems are to be addressed that need large boxes 
with $L>4$fm or small values of the lattice spacing with $a<0.05$fm, the 
computing expense will again be beyond present capabilities. 
Therefore, there is still the need for further developing algorithms and 
machines for lattice QCD.

Whether simulations are performed directly at the physical point or whether
chiral perturbation theory will be used to extrapolate to the physical point is 
a decision left to the particular collaboration performing such simulations. 
It is my belief, however, that we need both approaches and that we should 
understand the mass dependence of physical observables. There is a number of 
examples, e.g., moments of parton distribution functions, where the present results
at about $m_\mathrm{PS}=300$MeV are still pretty far away from the experimental value and 
it will be very interesting to see how the approach to the physical point 
is realized,
as this can provide a valuable insight into the physics of the considered problem.
In addition, precise determinations of the low energy constants 
of chiral perturbation theory from the mass dependence of physical 
observables will be one of the main accomplishments of lattice QCD. 

\section{Universality}

A demonstration of universality of lattice QCD, i.e. showing that different 
lattice fermion 
formulations give consistent continuum limit values for physical observables,
is, in my opinion, 
a crucial goal. 
Basically all present formulations of lattice QCD have some kinds of conceptual 
weaknesses (or are too expensive to simulate) leading to different kind of 
systematic effects which will (hopefully) disappear in the 
continuum limit. Checking that alternative lattice fermion 
formulations give consistent 
results in the continuum limit 
--and thus demonstrating universality-- is hence of utmost importance. 

\begin{figure}[t]
\centering
{\includegraphics[width=0.75\linewidth]{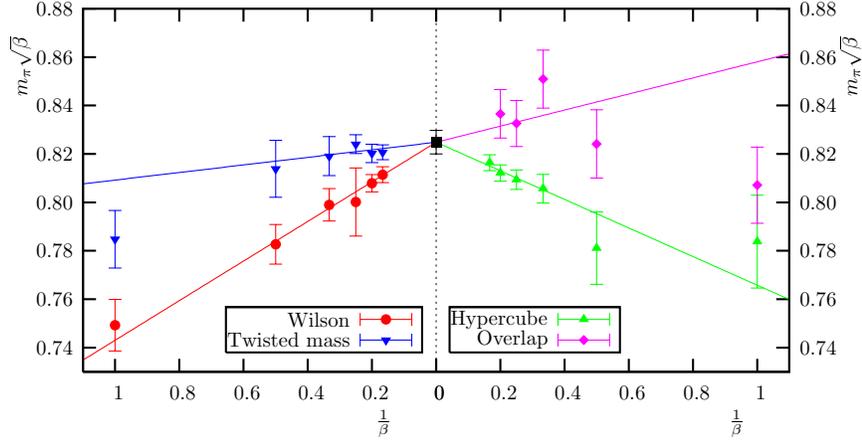}}
\caption{Schwinger model results for the lightest pseudo scalar
particle mass $\sqrt{\beta}M_\pi$ as a function of $a^2=1/\beta$. 
The continuum limit scaling is shown for Wilson, maximally twisted mass, 
hypercube and overlap fermions for a fixed value of the quark mass. 
The common continuum limit value for all these kind of lattice fermions 
demonstrates universality for this model.}  
\label{fig:schwinger}
\end{figure}

Let me illustrate this point with the example of the Schwinger model taken 
from ref.~\cite{Christian:2005yp}. In 
fig.~\ref{fig:schwinger} the continuum limit extrapolation of the mass of the 
lightest pseudo scalar particle, denoted here as $M_\pi$, 
in the Schwinger model is shown.
In this super-renormalizable model the coupling $\beta\propto 1/a^2$ can be used
as scaling variable and $M_\pi\sqrt{\beta}$ has a well defined continuum 
limit for a fixed physical quark mass, i.e. $m_\mathrm{quark}\sqrt{\beta}$ fixed. 
The graph in fig.~\ref{fig:schwinger} 
shows an example of the continuum limit for one choice of a fixed quark mass
using Wilson \cite{Wilson:1974sk}, maximally twisted mass \cite{Frezzotti:2003ni}, 
hypercube \cite{Bietenholz:1995cy} and overlap fermions \cite{Neuberger:1997fp}. 
Taken aside the overlap fermion simulations which have too large errors 
to be really conclusive, 
all formulations show the expected $a^2$ scaling behaviour and converge to the same
continuum limit value thus demonstrating nicely universality.

In my opinion, it would be most important to have analogous 
graphs for various quantities in case of lattice QCD demonstrating
convincingly 
that we can obtain 
consistent results in the continuum limit from various formulations
of lattice QCD. Unfortunately, we are not yet
in the position to show such a graph. On the contrary, we have even examples
where discrepancies seem to be visible when the continuum limit is
taken.
Let me discuss the situation 
here at the examples of the nucleon mass and the pseudo scalar decay constant. 

\subsection{Nucleon mass} 

For the following discussion, I will use $r_0$ \cite{Sommer:1993ce} 
as a scaling variable. This choice is motivated by the fact that here I am 
not interested in direct physical values in terms of MeV but only in the 
scaling behaviour. In addition, determining $r_0$ is by now a standard and well 
understood procedure \cite{Necco:2001xg} and which is used by many groups. It avoids the 
difficulty of using  
the lattice spacing itself which is often determined from different 
observables in the various collaborations thus leading possibly to 
large systematic effects.

\begin{figure}[t]
  \subfigure[Scatter graph of the nucleon mass as function 
of the pseudo scalar mass squared using $r_0$ to set the scale. 
Data are taken from maximally twisted mass fermions, 
ref.~\cite{Alexandrou:2008tn} (ETMC), 
rooted staggered fermions, ref.~\cite{Aubin:2004wf,Bernard:2007ux} (MILC), 
domain wall fermions, ref.~\cite{Allton:2008pn} (RBC-UKQCD), 
non-perturbatively improved Wilson fermions, 
ref.~\cite{Gockeler:2007rx} (QCDSF-UKQCD) 
and ref.~\cite{Aoki:2008sm} (PACS-CS), 
overlap fermions, ref.~\cite{Noaki:2008gx,Ohki:2008ff}
and domain wall fermions on rooted staggered
sea quarks, ref.~\cite{WalkerLoud:2008bp} (LHP).
A number of values presented in the graph
are from private communications.
  \label{fig:nucleon1}]
  {\includegraphics[width=0.55\linewidth]{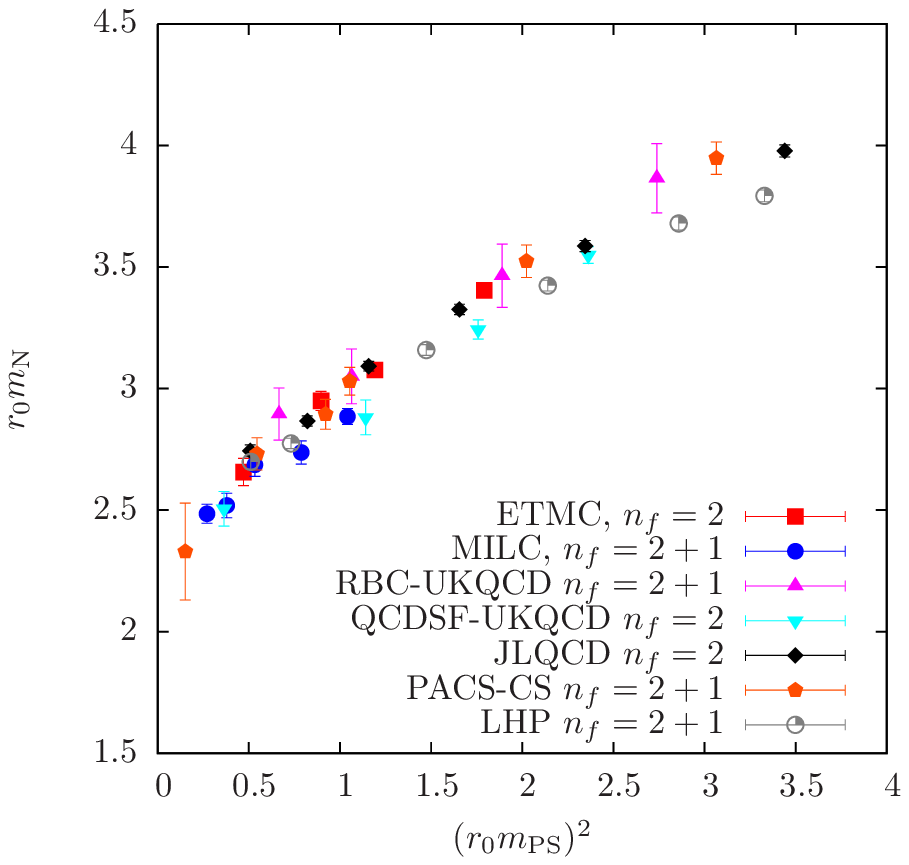}}\quad
  \subfigure[The nucleon mass as a function of the lattice 
spacing squared at fixed values of $r_0m_\mathrm{PS}$. 
For explanations what kind of fermions is used, see the left panel 
of the graph.
\label{fig:nucleon2}]
  {\includegraphics[width=0.55\linewidth]{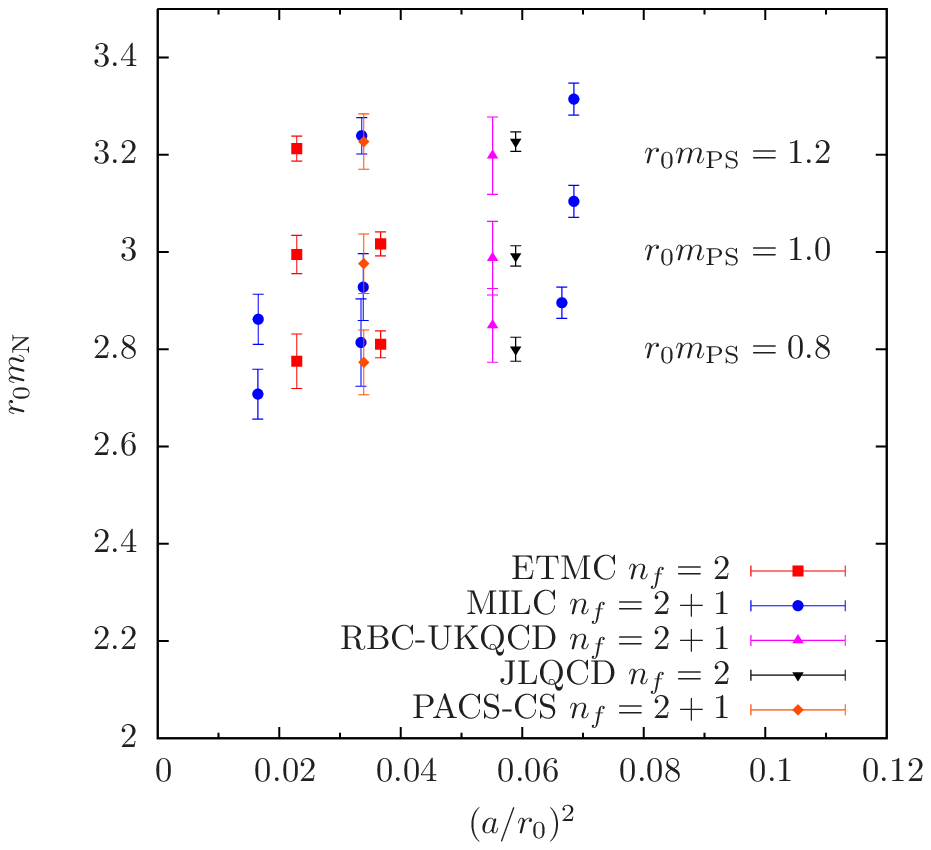}}
  \caption{The scaling behaviour of the nucleon mass.}
  \label{fig:nucleon}
\end{figure}

In the following, an attempt is made to show the continuum limit 
scaling for the nucleon mass 
$r_0M_\mathrm{nucleon}$ at fixed pseudo scalar masses 
$r_0m_\mathrm{PS}=0.8,1.0,1.2$.
Let me start with a compilation graph, fig.~\ref{fig:nucleon}, showing 
$r_0M_\mathrm{nucleon}$ versus $(r_0m_\mathrm{PS})^2$ as evaluated from 
a number of collaborations using Wilson, twisted mass, staggered, domain 
wall and overlap fermions, see the figure caption for corresponding
references.
The overall impression in this graph is a nice consistency of all the results 
and a rough scaling behaviour since all results fall into a rather narrow 
band. Note that in this graph results from $N_f=2$ and $N_f=2+1$ flavours 
of quarks are mixed. Of course, it is not too surprising that for the 
nucleon mass there is no big effect of having a dynamical 
strange quark.

The scaling behaviour is shown in more detail in fig.~\ref{fig:nucleon2}
where the nucleon mass
is plotted as a function of $(a/r_0)^2$ for three values of $r_0m_\mathrm{PS}$. 
The data follow basically the expected $a^2$ behaviour and are consistent with 
each other. 

In summary, for the nucleon sector the scaling properties look 
promising and with results at more values of the lattice spacing,  
as will be obtained in the near future, a detailed scaling 
comparison can be performed. 

\subsection{The pseudo scalar decay constant} 

\begin{figure}[t]
  \subfigure[Scatter graph of the pseudo scalar decay constant as function
of the pseudo scalar mass squared using $r_0$ to set the scale.
Data are taken from maximally twisted mass fermions,
ref.~\cite{Boucaud:2007uk,Urbach:2007rt} (ETMC),
rooted staggered fermions, ref.~\cite{Bernard:2007ps} (MILC),
domain wall fermions, ref.~\cite{Allton:2008pn} (RBC-UKQCD),
non-perturbatively improved Wilson fermions, 
ref.~\cite{Gockeler:2006vi} (QCDSF-UKQCD) and 
ref.~\cite{DelDebbio:2007pz} (CERN) 
and overlap fermions \cite{Noaki:2007es} (JLQCD). 
A number of values
are taken from private communications of the various collaborations.
  \label{fig:fps1}]
  {\includegraphics[width=0.55\linewidth]{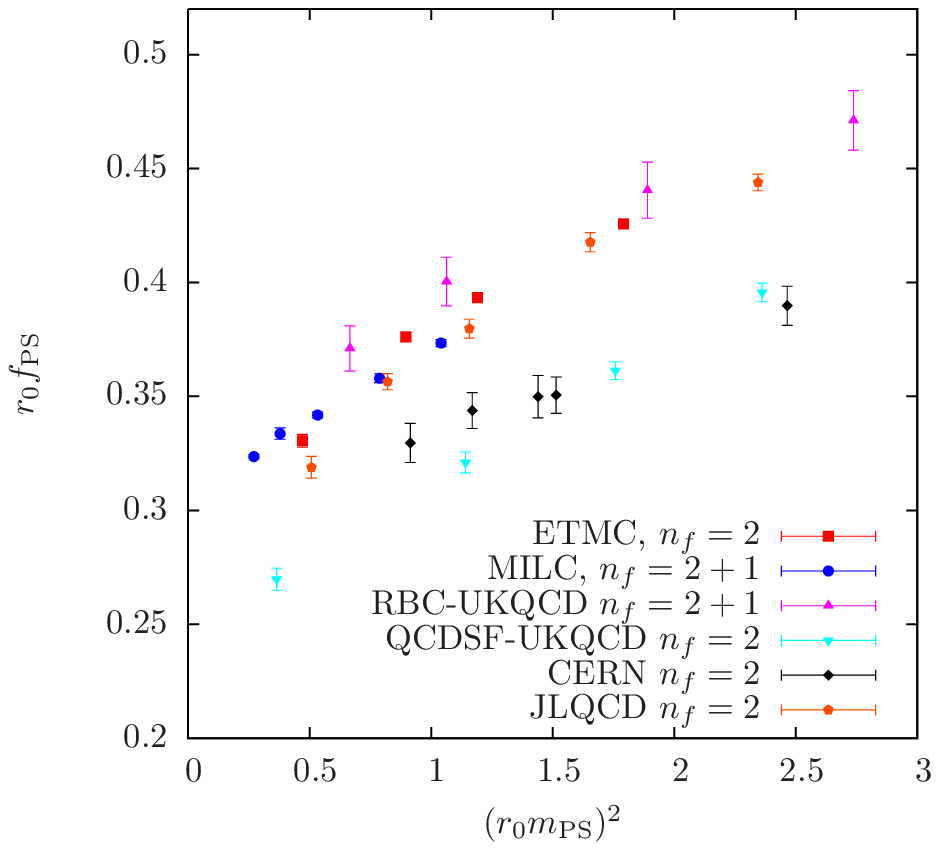}}\quad
  \subfigure[The scaling in the lattice spacing of the pseudo scalar 
decay constant as function of the lattice spacing squared at fixed
$r_0m_\mathrm{PS}$. Only those formulations of lattice QCD are taken 
for which no explicit renormalization is necessary. \label{fig:fps2}]
  {\includegraphics[width=0.55\linewidth]{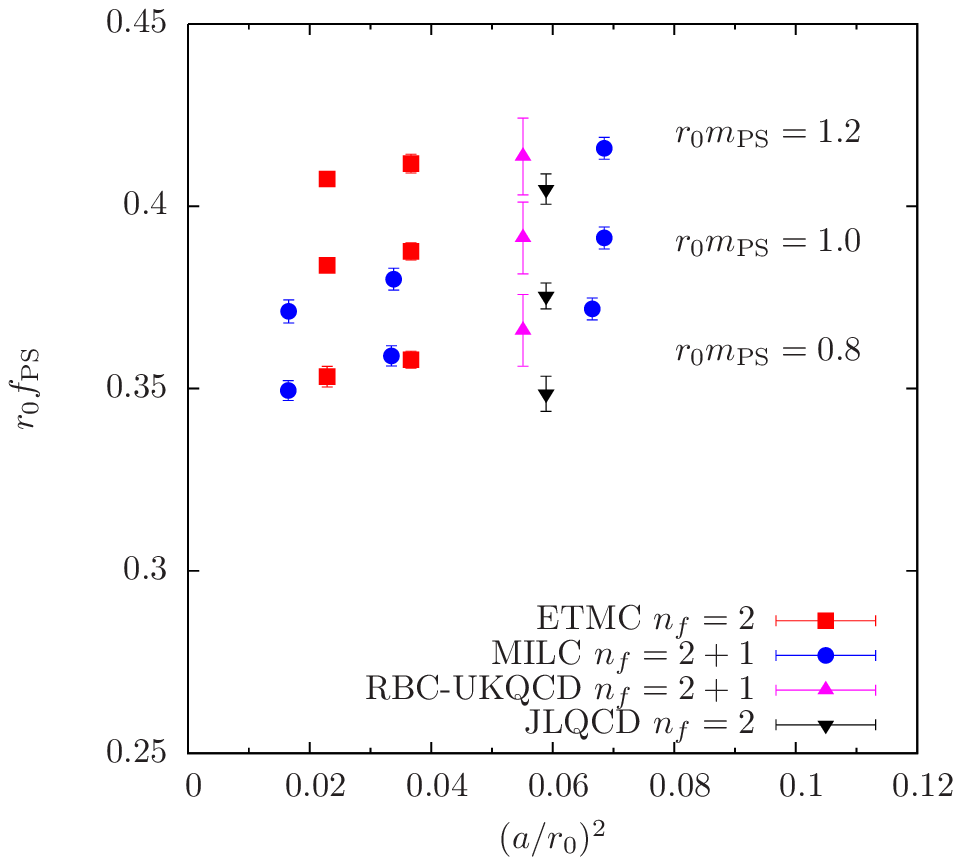}}
  \caption{Lattice spacing scaling of the pseudo scalar decay constant.}
  \label{fig:fps}
\end{figure}

In fig.~\ref{fig:fps1} a compilation of various results for $r_0f_\mathrm{PS}$ versus
$(r_0m_\mathrm{PS})^2$ is shown. This graph is very surprising and, at least to me, 
rather scary. In contrast to the corresponding compilation graph for the 
nucleon mass in fig.~\ref{fig:nucleon}, the data for $r_0f_\mathrm{PS}$ scatter 
a lot and do not show a common scaling behaviour. 

The cause of the apparent inconsistencies shown in 
fig.~\ref{fig:fps1} is not clear presently. 
One possible reason could be that for a number 
of formulations, such as Wilson fermions a renormalization of $f_\mathrm{PS}$ 
is required. I therefore show in fig.~\ref{fig:fps2}
the scaling of $r_0f_\mathrm{PS}$ for only 
those lattice fermion formulations for which an explicit renormalization 
is not required, i.e. staggered, maximally twisted mass and overlap fermions. 
Here the situation looks indeed better and a rough consistency among these 
results can be seen. 

Of course, this does not mean that it is indeed the renormalization of 
$f_\mathrm{PS}$ that is behind the very visible differences for 
$f_\mathrm{PS}$ from different fermions. This is in particular so, 
since precise non-perturbative computations of $Z_A$ 
are available \cite{DellaMorte:2008xb}. Other causes could be the 
values of $r_0$ used in the comparison and finite size effects can be 
significant in $f_\mathrm{PS}$ as is discussed also below, 
although in the analysis used here the data for $f_\mathrm{PS}$ were 
finite size corrected.
Another possibility is that  
the values of the lattice spacing might be still too coarse.             
Finally, it might be that we see a problem with $f_\mathrm{PS}$ and 
seemingly not with $m_\mathrm{N}$ because the data for $f_\mathrm{PS}$ 
are much more precise and that only such an accuracy can reveal  
lattice spacing artefacts, i.e. that there might still be large 
$\mathrm{O}(a^2)$ effects.

Which of the above mentioned possibilities will turn out to be the culprit 
in the end, or whether there is a completely different cause, is not 
possible to say at the moment. However, I think that the lattice community 
must investigate this issue in the future. For me, a clarification 
of the problem with $f_\mathrm{PS}$ should be high position on 
the priority list.

\section{The actions}

In the introductory section I have given two examples of precision 
{\em continuum} calculations coming from lattice QCD simulations, 
namely the classical benchmark computation of the baryon 
spectrum and the accurate determination of low energy constants
of chiral perturbation theory.

These nice results, however, 
do not mean that we have lattice QCD fully under control yet. 
A striking example is the lattice spacing scaling of the pseudo scalar
decay constant discussed above.  
As argued already, a most important 
point is therefore the verification of universality. The lattice formulations of 
QCD used today all have their shortcomings each leading to a number of 
systematic effects and only reaching consistent continuum results from 
alternative formulations will show that such systematic errors are 
under control. 
Let us go shortly through a number of different formulations of lattice fermions 
and discuss their shortcomings.

\subsection{Wilson fermions}

Wilson fermions \cite{Wilson:1974sk} with improvement terms 
\cite{Sheikholeslami:1985ij} and non-perturbative 
improvement \cite{Jansen:1998mx,Aoki:2005et} are used widely in lattice calculations. 
Their major drawback 
--besides the demanding computation of the non-perturbative operator improvement-- 
is the explicit breaking of chiral symmetry at non-vanishing 
values of the lattice spacing. In the past, when using the quenched approximation, 
one of the consequences was the 
appearance of unphysical, small eigenvalues of the Wilson-Dirac operator.  

With modern simulations of lattice QCD employing the quarks as dynamical degrees
of freedom, it turns out, however, that these small eigenmodes do not 
appear even when much smaller values of the pseudo scalar mass 
are simulated than it was possible in the quenched approximation. 
In fact, in ref.~\cite{DelDebbio:2007pz} a stability criterion has been developed, 

\begin{equation}
m_\mathrm{PS} L \ge \sqrt{3\sqrt{2}aB/Z}
\label{stability}
\end{equation}

\noindent providing a bound on the pseudo scalar mass 
down to which stable simulations can be performed. 
In eq.~(\ref{stability}), $B$ is a low energy constant of chiral 
perturbation theory related to the scalar condensate and $Z$ the quark mass 
renormalization constant. This bound derives from the observation that there is
a spectral gap and from the demand that this gap is, say, 
three times larger
than the width of the corresponding eigenvalue distribution. 

In recent years, another feature of Wilson type fermions has been observed. 
When approaching, for sufficiently large values of the gauge coupling 
$\beta$,
the chiral limit at zero quark mass, 
a rather strong first order phase transition occurs. 
This phenomenon is a remnant of the continuum first order phase transition when 
changing the quark mass from positive to negative values. 

The lattice distorted first order phase transition has been anticipated already 
in ref.~\cite{Sharpe:1998xm}. 
First signs of such a phase transition have been seen in 
refs.~\cite{Blum:1994eh,Aoki:2001xq,Aoki:2004iq,Jansen:2003nt} 
and thorough numerical investigations have been performed in 
refs.~\cite{Farchioni:2004us,Farchioni:2004ma,Farchioni:2004fs,Farchioni:2005ec,Farchioni:2005bh,Farchioni:2005tu} in the twisted mass formulation. 
These numerical findings are in accord with results from 
chiral perturbation theory, see refs.~\cite{Sharpe:1998xm,Sharpe:2004ny,Aoki:2004ta,Sharpe:2004ps,Munster:2003ba,Munster:2004am,Munster:2004wt,Scorzato:2004da}, and a complete 
picture resulted from these works.
As an aside, we also mention that at small values of $\beta$ an 
Aoki phase \cite{Aoki:1986ua} with a spontaneous breaking of parity appears
\cite{Sternbeck:2003gy,Ilgenfritz:2003gw}. 

\begin{figure}[t]
  \subfigure[Eigenvalue distribution of the (Hermitean) Wilson-Dirac operator
   at $\beta=5.2$ on a $16^4$ lattice. The median $\bar{\mu}$ 
   of this distribution is $\bar{\mu}=0.0103$ while its width $\sigma$ 
   is $\sigma=0.0013$ thus obeying the bound given in of eq.~(\protect\ref{stability}).
  \label{fig:stability1}]
  {\includegraphics[width=0.55\linewidth]{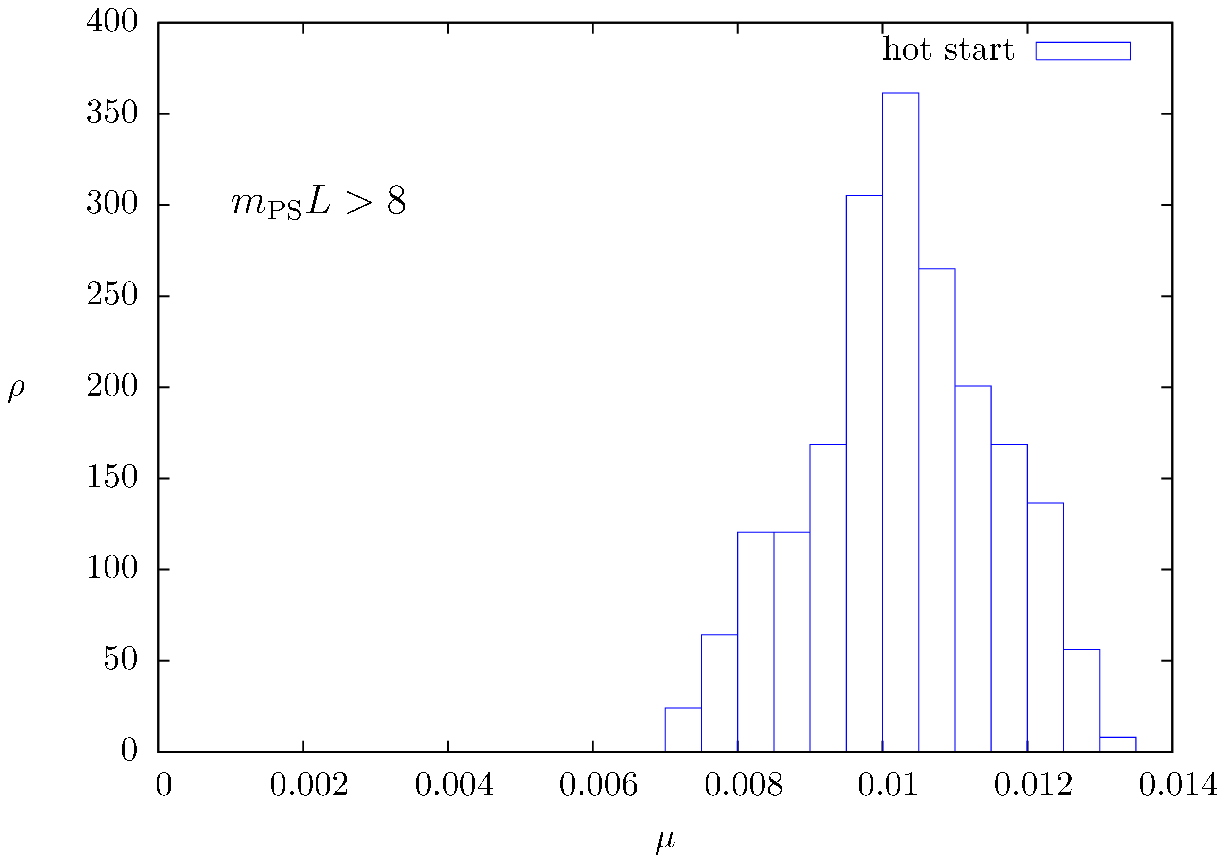}}\quad
  \subfigure[The Monte Carlo time history of the plaquette starting 
   from hot and cold configurations. The parameters are the same as 
   those in the left panel. A clear metastable behaviour is observed. 
   \label{fig:stability2}]
  {\includegraphics[width=0.55\linewidth]{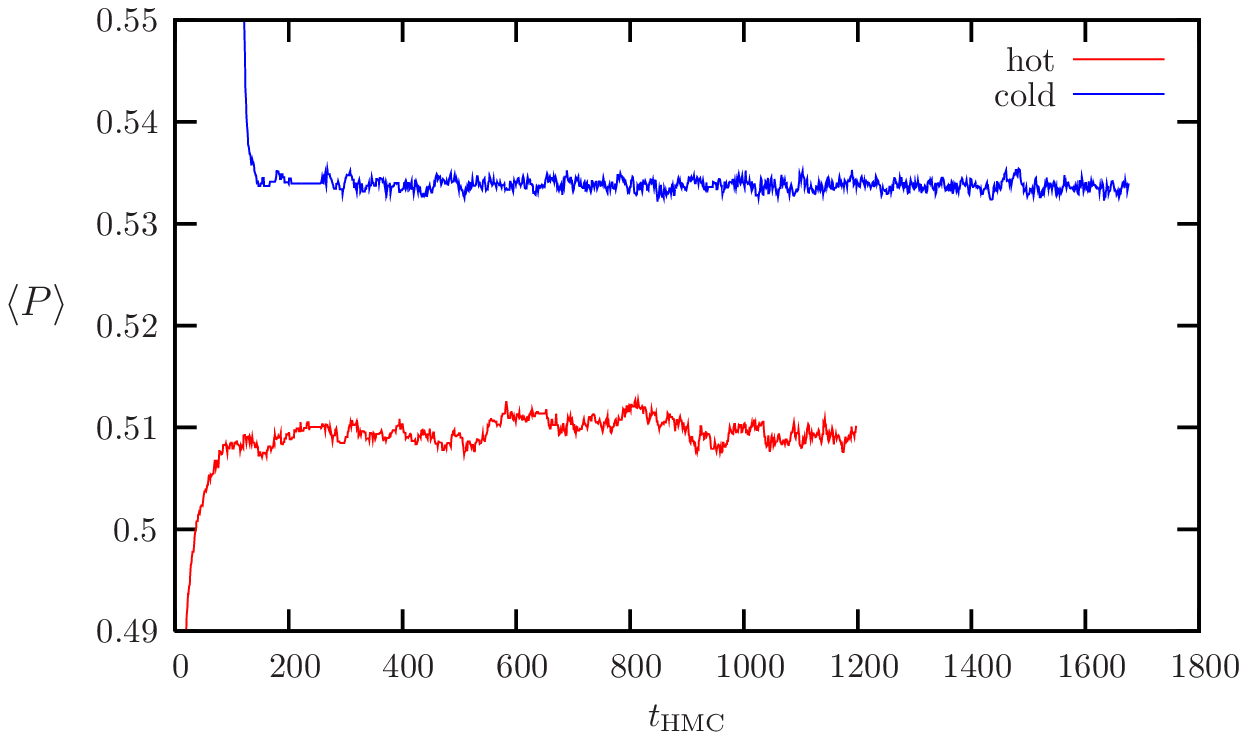}}
  \caption{First order phase transition and stability \protect\cite{Urbachunpublished}.}
  \label{fig:stability}
\end{figure}

The strength of the first order phase transition strongly depends on the value
of the lattice spacing and of the twisted mass 
used in the simulation. It is clearly visible at rather coarse 
lattice spacings and can there even invalidate the stability criterion discussed above. 
This is demonstrated in fig.~\ref{fig:stability} \cite{Urbachunpublished}. 
From the width of the eigenvalue distribution 
shown in fig.~\ref{fig:stability1} in the left panel one would 
conclude that the simulations are stable 
and safe. However, in the right panel, fig.~\ref{fig:stability2},
a metastable behaviour of the simulation
is observed when starting with hot and cold configurations. Thus, this simulation
point, although fulfilling the stability criterion, suffers from metastable 
behaviour. 
Let me remark that the value of the lattice spacing used in this investigation 
has been large, $a>0.1$fm.

Although with decreasing lattice spacing for fixed twisted mass (zero or non-zero)
the effects of the first order 
phase transition gets weaker and the stability criterion may become 
more relevant, I still think that 
it is not sufficient to {\em only} check the median and 
the width of the eigenvalue
distribution but to also check for the existence of a possible first order phase
transition. Checks on both the existence of meta stabilities and the 
stability criterion from the eigenvalue distribution at the actual 
simulations points have become 
routine for a number of collaborations already. 
As a result of such checks, Wilson fermions are in the 
fortunate situation that by
respecting the bound in eq.~(\ref{stability}) and avoiding  
meta stabilities, e.g. by going to sufficiently small lattice spacings,  
simulations can be expected to be performed and controlled even when 
applied at pseudo scalar masses as small as 200MeV or even at the physical 
point.

\subsection{Staggered fermions}

The staggered fermion community is comprised of mainly the MILC  
collaboration \cite{MILC}, at least as far as the generation of gauge field
configurations is concerned. MILC has by now produced a large and impressive 
set of configurations 
with dynamical up and down as well as strange quark degrees of freedom. 
These configurations are also uploaded to the International Lattice Data 
Grid (ILDG). (See ref.~\cite{Yoshie2008} for a recent overview on ILDG.)
MILC has produced these configurations in a project which is ongoing 
now for many years and has produced configurations at 
small values of the lattice spacing of $a=0.06$fm and correspondingly 
large lattice sizes with $64^3\cdot 144$ number of  
lattice points to obtain a reasonable box size in physical units. 
There are furthermore plans for future runs with a lattice spacing of $a=0.045$fm. 

Therefore, the question whether this approach to lattice QCD has a conceptual flaw
when taking the fourth root is of the greatest importance. The last years
have seen many discussions on the issue, see 
refs.~\cite{Jansen:2003nt,Adams:2004wp,Durr:2005ax,Sharpe:2006re,Kronfeld:2007ek} for reviews
on the subject. The locality of rooted staggered fermions is addressed 
in Shamir's 
work \cite{Shamir:2004zc,Shamir:2006nj}. A controversial and still 
ongoing debate can be found in 
refs.~\cite{Creutz:2006wv,Creutz:2007pr,Creutz:2007rk,Creutz:2007yr,Creutz:2008hx,Creutz:2008kb} and 
refs.~\cite{Golterman:2006rw,Bernard:2006vv,Bernard:2006ee,Bernard:2006qt,Bernard:2007eh,Bernard:2007ma,Bernard:2008gr}. It is not the aim of this contribution 
to enter this debate or to even judge between the opponents. 
However, the picture that 
emerges --at least to my understanding-- can be summarized 
in two scenarios.

In scenario one, the {\em practitioner scenario}, we do not insist that we reach the 
{\em chiral limit} at zero quark mass for non-zero values of the 
lattice spacing. Rather, we follow a procedure
to stop at some threshold quark mass for a given value of the lattice spacing and then
first perform the continuum limit and only afterwards the extrapolation
to the physical point. A discussion using staggered 
chiral perturbation theory to obtain bounds on such threshold
quark mass values  
can be found in refs.~\cite{Bernard:2006zw,Bernard:2007ma}. 
A summary of these results is that for applying {\em continuum} chiral perturbation
theory the taste splitting of staggered fermions has to be much smaller than 
the lightest pseudo scalar mass. If 
instead staggered chiral perturbation theory is applied, 
the taste splitting can be at the order of the lightest pseudo scalar mass, since 
the taste breaking effects can then be taken into account. 
For example \cite{bernardpriv}, at a lattice 
spacing of $a\approx 0.06$fm the lightest pseudo scalar mass simulated is about
$m_\mathrm{PS}=220$MeV which is three times larger than the observed 
taste splitting. For $a\approx 0.125$fm the lightest pseudo scalar mass 
of $250$MeV 
is about the order of the taste splitting and it would thus make not much sense
to simulate even smaller masses.
More general bounds on the quark mass that follow from the 
locality considerations of rooted 
staggered fermions can be found in \cite{Shamir:2006nj}.
A nice discussion of the question of 
interchanging continuum and chiral limits is given in 
\cite{Durr:2004ta} for the case of the 1-flavour Schwinger model.

In scenario two, the {\em theorist scenario},
we want to explore the behaviour of staggered fermions with the 
fourth root trick at or very close to the chiral point. This could 
reveal some non-perturbative effects of the fourth root trick 
(e.g. related to the 't Hooft vertex) which 
could eventually lead to a failure of this approach to lattice QCD. However, 
when respecting the bounds on the quark mass discussed above, 
the possible difficulties of rooted staggered fermions in the chiral 
limit may not
affect the results obtained following scenario one. 
Possible quantities to explore the extreme chiral regime are
those related to instanton physics.
In my opinion, the exploration of the chiral limit for staggered fermions is 
of theoretical importance and further scientific discussions, beyond 
the literature given above, on the topic are 
certainly welcome.
An investigation on this topic can be found in \cite{Adams:2008db}.

Another disturbing observation about present staggered fermion simulations 
is the fact that
for the very large lattice simulations an inexact Hybrid Monte Carlo algorithm 
is used. The inexactness comes from the fact 
that no accept/reject step is applied at the end of a 
molecular dynamics 
trajectory. Although there are some arguments and investigations that this might 
be a harmless procedure \cite{toussaint:priv}, doubts are legitimate and 
re-introducing the accept/reject 
step would certainly enlarge the trust in the staggered fermion simulations. 

\subsection{Twisted mass fermions}

Twisted mass fermions at maximal twist \cite{Frezzotti:2003ni,Frezzotti:2004wz}
have by now proved to be a
practical and successful tool for performing lattice QCD simulations, see 
e.g. 
refs.~\cite{Boucaud:2007uk,Blossier:2007vv,Cichy:2008gk,Boucaud:2008xu,Alexandrou:2008tn,Jansen:2008wv,Shindler:2007vp} and contributions to this conference
\cite{Jansen:2008ht,Lopez:2008ns,Jansen:2008ru,Dimopoulos:2008ee,Dimopoulos:2008sy,Baron:2008xa}. 
The expected $\mathrm{O}(a)$-improvement \cite{Frezzotti:2003ni} 
has been demonstrated for many 
observables by now in the quenched approximation
\cite{Jansen:2003ir,Jansen:2005gf,Jansen:2005kk,Abdel-Rehim:2005gz} as well 
as employing dynamical 
quarks \cite{Boucaud:2007uk,Urbach:2007rt,Dimopoulos:2007qy,Boucaud:2008xu}.
In particular,  
it was shown that stable simulations down to 
pseudo scalar masses of about $m_\mathrm{PS}\approx 260$MeV are possible. 

\begin{figure}[t]
  \subfigure[The difference of the charged and neutral pseudo scalar 
   mass as a
  function of $a^2$ in the twisted mass formulation of lattice QCD at 
  different values of the charged pseudo scalar mass. A large 
  $\mathrm{O}(a^2)$ lattice artefact is observed.
  \label{fig:tm1}]
  {\includegraphics[width=0.55\linewidth]{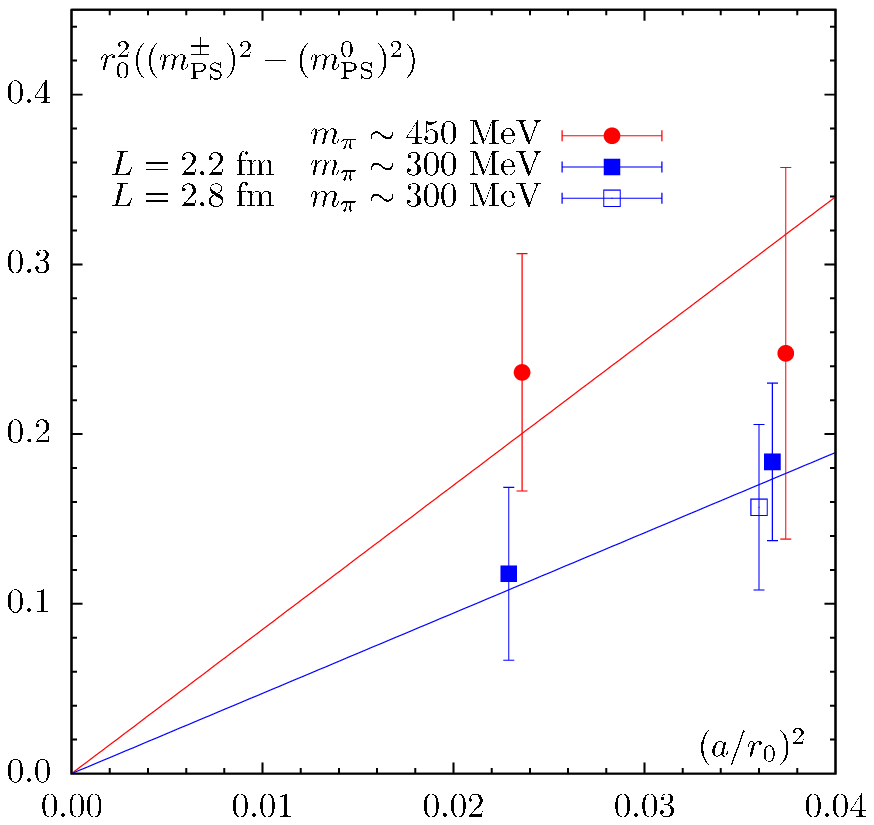}}\quad
  \subfigure[The scaling behaviour of the charged pseudo scalar mass at 
   fixed values of the pseudo scalar decay constant. The scaling behaviour 
   is basically flat in $a^2$ demonstrating that the cutoff effect 
   in the left panel originates solely from the neutral pseudo scalar 
   mass.   \label{fig:tm2}]
  {\includegraphics[width=0.55\linewidth]{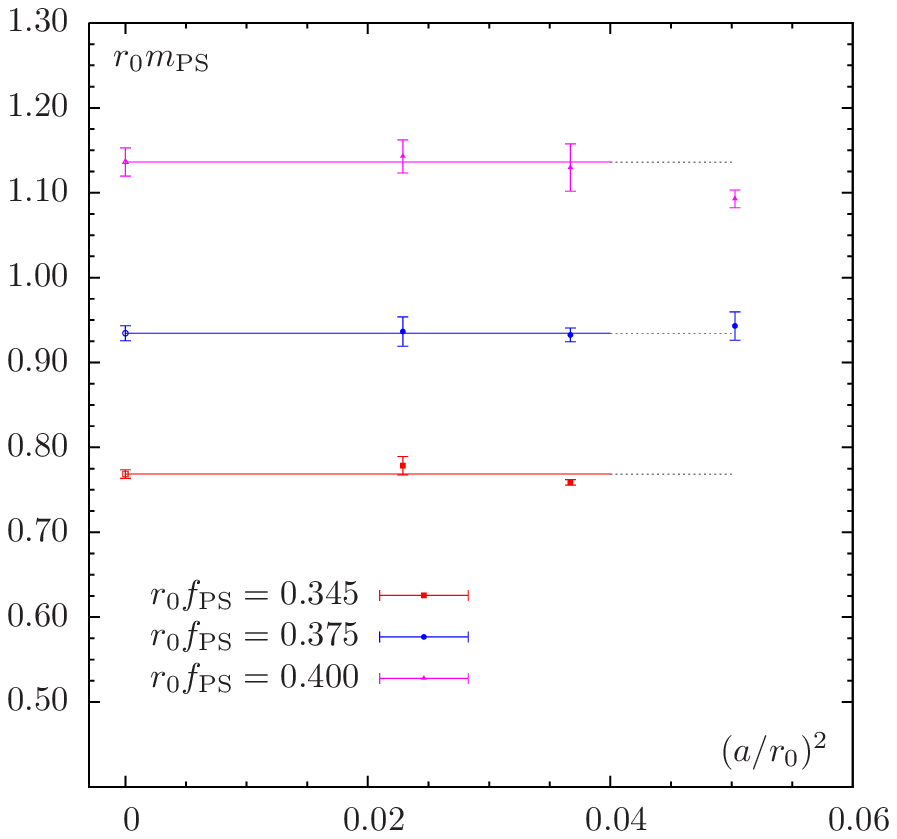}}
  \caption{Isospin violations for twisted mass fermions at the example 
   of the charged and neutral pseudo scalar masses.}
  \label{fig:tm}
\end{figure}

Twisted mass fermions share with standard Wilson fermion the drawback of 
breaking  
chiral symmetry at any non-zero lattice spacing.
An additional major drawback of twisted mass fermions is the explicit violation of 
isospin symmetry at non-zero values of the lattice spacing. From the 
simulations by the European twisted mass collaboration (ETMC) there 
are two basic observations concerning this 
lattice artefact. The first is that the isospin breaking, although 
consistent with the expected $\mathrm{O}(a^2)$ scaling, is large when the 
mass difference of the charged and the neutral pseudo scalar mass 
is considered as can be seen in fig.~\ref{fig:tm1}. Note that for the computation 
of the neutral pseudo scalar mass disconnected diagrams need 
to be taken into account. In contrast, the
scaling behaviour of the charged pseudo scalar mass is very flat 
showing almost no lattice artefacts as demonstrated in fig.~\ref{fig:tm2}. 
Thus, the large lattice artefact seen in fig.~\ref{fig:tm1} must be due
to the neutral pseudo scalar mass alone. 

A second observation is that other quantities seem not to be affected 
by the isospin violation, as can be seen from table~\ref{tab:isospin}. 
There the relative difference of charged and neutral quantities, 
$R_O=(O^\pm-O^0)/O^\pm$ turns out to be compatible with zero, at least 
within the errors. 

\begin{table}
\begin{center}
      \begin{tabular*}{0.5\linewidth}{@{\extracolsep{\fill}}lccr}
        \hline\hline
        $\Bigl.\Bigr.$ & $\beta$ & $a\mu_q$ & $R_O~~~~~~$ \\
        \hline
        $\Bigl.\Bigr.$ $af_\mathrm{PS}$ & $3.90$ & $0.004$ & $0.04(06)$ \\
        & $4.05$ & $0.003$ & $-0.03(06)$ \\
        \hline
        $\Bigl.\Bigr.$ $am_\mathrm{V}$  & $3.90$ & $0.004$ & $0.02(07)$ \\
        & $4.05$ & $0.003$ & $-0.10(11)$ \\
        \hline
        $\Bigl.\Bigr.$ $af_\mathrm{V}$  & $3.90$ & $0.004$ & $-0.07(18)$ \\
        & $4.05$ & $0.003$ & $-0.31(29)$ \\
        \hline
        $\Bigl.\Bigr.$ $am_\Delta$      & $3.90$ & $0.004$ & $0.022(29)$  \\
        & $4.05$ & $0.003$ & $-0.004(45)$ \\
        \hline\hline
      \end{tabular*}
\end{center}
\caption{Examples of relative differences between charged and neutral 
operator expectation values, $R_O=(O^\pm-O^0)/O^\pm$, 
measuring the isospin breaking effects 
in twisted mass lattice QCD.}
\label{tab:isospin}
\end{table}

The two observations described above find an interpretation in terms of the 
Symanzik effective theory analysis \cite{Frezzotti:2007qv}. 
In particular, there it can be 
shown that the charged pseudo scalar mass receives only $\mathrm{O}(a^2m_\pi^2,a^4)$ 
corrections while the neutral pseudo scalar mass has corrections 
at $\mathrm{O}(a^2)$. This explains the scaling behaviour 
of the masses shown in fig.~\ref{fig:tm}. 
The results listed in table~\ref{tab:isospin} must then correspondingly 
be interpreted that in these quantities the neutral pseudo scalar mass 
(or related quantities) do not play a dominant role.

Whether the Symanzik type analysis provides the correct interpretation of the 
numerically obtained results or whether other interpretations are possible and, 
maybe, more applicable is presently being investigated by ETMC. In any case, there 
exists no general argument, whether or not large isospin breaking effects 
can appear in certain quantities. 
Although there exist indications that large isospin breaking effects may only appear in
certain observables (like the neutral pion mass), this issue must be studied 
carefully on a case by case basis by any group
employing twisted mass fermions.

\subsection{Smearing}

Many simulations use nowadays some method 
of smearing of the links \cite{Morningstar:2003gk,Durr:2007cy,Hasenfratz:2007rf}
that enter the 
lattice-Dirac operator. 
This procedure has the advantage to smooth out the configurations
seen by the lattice Dirac operator. As a consequence, the condition number can be 
reduced and also the gauge field fluctuations are suppressed leading to 
possibly faster and more stable 
simulations compared to the case when no smearing is employed. 
In addition, also the effects of the first order phase transition 
mentioned above seem to be diminished \cite{Jansen:2007sr}. 

An open question is of course, to what extent smearing should be used. 
Performing only moderate smearing as done in 
e.g. refs.~\cite{Horsley:2008ap,Schaefer:2007dc}  
will presumably not affect
the simulations much. However, already one level of stout smearing 
\cite{Morningstar:2003gk} 
leads in perturbation theory \cite{Capitani:2006ni} to values for
the renormalization constants and improvement coefficients that are close
to their tree-level values when smearing is employed \cite{Horsley:2008ap,Horsley:2008fw}.
This holds at least for certain values of the smearing parameter $\rho\approx 0.1$.

When many levels of 
smearing are performed as used in ref.~\cite{Durr:2008rw}, 
there is the danger that uncontrolled systematic 
effects emerge as the Dirac operator may become quite non-local. 
In the simulation,  
which revealed the baryon spectrum shown in fig.~\ref{fig:baryonspectrum},  
6-levels of stout smearing 
had been used \cite{bmw2008}. Many concerns about a possible alteration of the 
short distance behaviour
of physical quantities have been put forward by this rather high level of smearing and 
a suspicion that the action is too non-local has been raised.
The BMW collaboration themself has performed a locality test  
following the principle idea of ref.~\cite{Hernandez:1998et}. Note that 
a locality test of the Dirac-operator itself will not reveal 
any non-local effects since it anyway acts on nearest neighbours only.

Therefore, the quantity investigated has been  
the response of the Dirac operator
$D(x,y)$ 
with respect to a gauge link variation 
$\|\partial D(x,y)/\partial U_\mu(x+z)\|$ 
as a function of the distance $z/a$. 
With a smearing parameter $\rho$ chosen to be well below 
one, it can be expected that smearing effects decay like $\rho^n$ and hence
the effects of smearing vanish rapidly for increasing distances. 
However, it is important to realize
that there is a high degree of degeneracy of lattice 
points at large distances which become relevant through the 
smearing procedure. 
In addition, in most quantities the behaviour of the inverse fermion matrix
(propagator) matters, not of the fermion matrix itself. 
This might strongly increase the effect of a 
high level smearing. Thus it is unclear what the net effect of smearing will be. 

\begin{figure}[t]
  \subfigure[The response of the Wilson Dirac operator on a variation 
   of a gauge field at a distance $z/a$. The graph illustrates that locality 
   is realized with an exponentially fast decrease of the norm  
    of $\|\partial D(x,y)/\partial U_\mu(x+z)\|$.
  \label{fig:locality}]
  {\includegraphics[width=0.43\linewidth]{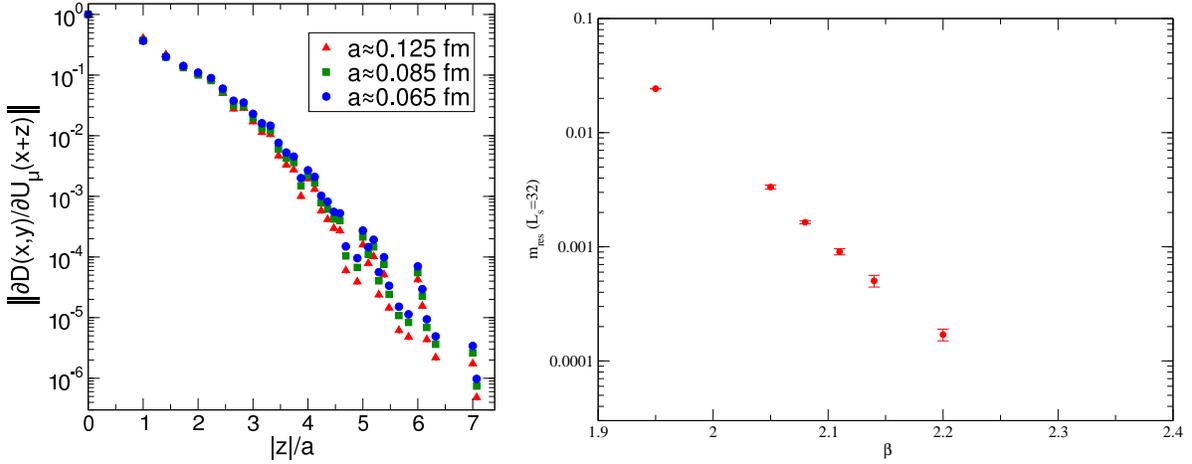}}\quad
  \subfigure[The residual mass $m_\mathrm{res}$ at fixed value of the extra dimension, 
   $L_s=32$, as a function of the gauge coupling $\beta$. Note the logarithmic scale 
   indicating the exponentially fast vanishing of $m_\mathrm{res}$. \label{fig:dw}]
  {\includegraphics[width=0.580\linewidth]{mresl32.eps}}
  \caption{Effects of smearing on the locality (left panel). 
   The behaviour of the residual mass as function of $\beta$ (right panel).}
  \label{fig:localitydw}
\end{figure}

The outcome of the locality test by the BMW collaboration is shown in 
fig.~\ref{fig:locality}. 
For this test three different values of the lattice spacing
were used as indicated in the graph.
The data demonstrates that there is an exponentially fast decay 
of the norm of the variation of the lattice Dirac operator with respect 
to the gauge field $U_\mu(x+z)$ as a function of $z/a$. 
Thus, an action with 6-levels of stout smearing
still shows an exponential localization. 
In this respect, it is similar to the locality properties 
of overlap fermions. Therefore, although a strict transfer matrix 
is missing when high-levels of smearing are performed, 
the action can be considered as being local in the 
field theoretical sense. 
It might still be that certain short distance quantities, such as the scalar 
condensate, renormalization factors or the Coulomb part of the static potential 
are affected by smearing. But, so far there is no convincing evidence for such 
a distortion.
The positive --and, maybe, rather surprising-- outcome of the locality 
investigation of the BMW collaboration suggests that it would be very 
worthwhile to investigate high-level stout smearing further on and  
test or rule out possible conceptual shortcomings.

\subsection{Fermions with exact/approximate lattice chiral symmetry}

\vspace{0.5cm}
\noindent {\bf Domain wall fermions}
\vspace{0.5cm}

Domain wall fermions \cite{Kaplan:1992bt,Furman:1994ky} 
are {\em only} chiral invariant in the limit of an 
{\em infinite} extra dimension. They are theoretically on the same footing 
(see ref.~\cite{Kikukawa:2000bp} and references therein)
as overlap fermions \cite{Ginsparg:1981bj,Neuberger:1997fp,Neuberger:1998wv}. 
It is important to realize that 
truncating the number of slices in the extra dimension
is equivalent to reduce, e.g., the degree of the polynomial when constructing the overlap 
operator. In both ways chirally improved actions are obtained.  
However, chiral symmetry will be broken explicitly, the  
effects of which ought to be studied.
For domain wall fermions such investigations have been 
performed by the RBC-UKQCD collaborations 
in refs.~\cite{Allton:2007hx,Antonio:2008zz,Allton:2008pn}. 
The outcome of these investigations is that chirality breaking effects
can essentially be quantified by the size of the residual quark mass in 
relation to the quark mass values employed in the simulations.
When comparing the residual mass $m_\mathrm{res}$ 
with the sea and valence quark masses
in recent domain wall simulations, at a coarse value of the lattice spacing,
$m_\mathrm{res}$ is dangerously close to the sea quark mass 
and even bigger than the valence quark mass. However, for smaller 
lattice spacing the situation improves considerably. 
It is worth to stress that domain wall fermion simulations are 
not too much more expensive than Wilson-type fermion simulations as 
illustrated in fig.~\ref{fig:cost1}. In addition, algorithmic tricks such 
as inexact deflation or multigrid ideas can also be applied for domain wall 
fermions thus leading to possibly large improvement factors.

One interesting observation from recent domain wall simulations is the 
behaviour of the residual mass as a function of the lattice spacing. 
As fig.~\ref{fig:dw} shows, for a fixed value of the number of slices 
in the extra dimension $L_s$ the residual mass vanishes exponentially fast 
with decreasing lattice spacing. Since the residual mass is proportional 
to the eigenvalue density at zero eigenvalues, this means that at some value of 
$\beta$ the topological charge will not change anymore. 
A corresponding observation has been made by groups using the overlap operator
\cite{Hashimoto2008}.  
These findings are a consequence of the fact that at small enough values 
of the lattice spacing, the plaquette bound for the existence of a spectral
gap of the Wilson-Dirac operator of ref.~\cite{Hernandez:1998et} 
is satisfied leading to a spectral gap of the corresponding 
kernel Dirac operator and therefore no topology change can occur. 
This can lead to a severe conceptual problem for overlap or domain wall 
fermion simulations.
The spectral gap itself on the other hand is a consequence of the 
negative bare quark mass employed in the kernel operator. 
For standard Wilson-type fermions, the bare quark mass is 
on the other hand positive and hence the above arguments do not apply.
Of course, this does not exclude that also standard Wilson-type fermions 
can run into problems with topology changes at large values of 
$\beta$.

\vspace{0.5cm}
\noindent {\bf Overlap fermions}
\vspace{0.5cm}

The statement that the cost 
for overlap or domain wall fermions with {\em exact} lattice chiral symmetry 
is at least one order of magnitude larger than 
for Wilson or staggered fermions, is, unfortunately still true today 
(see e.g. ref.~\cite{Schaefer:2006bk}) although 
many developments and improvements have already been taken place. 
The reasons are the 
nested iterations in inverting the operator and the difficulty to tunnel 
between different topological charge sectors. 

Nevertheless, simulations on small lattices are performed nowadays and some
first results are emerging
\cite{DeGrand:2006nv,DeGrand:2007tm,Cundy2008,Cundy:2007dp,Cundy:2008zc}. 
However, it seems to me that chiral invariant 
simulations in lattice QCD are still a subject for the future\footnote{Note, however, 
that in chiral invariant Higgs-Yukawa like models which employ the 
overlap operator 
\cite{Gerhold:2007pj,Fodor:2007fn,Gerhold:2007gx,Gerhold2008} 
lattices with size $32^3\cdot 64$ are used already.}.
 
\vspace*{0.5cm}
\noindent {\bf Fixed topology simulations}
\vspace*{0.5cm}

As a solution to the topology tunneling problem of overlap simulations, the usage 
of topology fixing actions has been put forward since these actions avoid 
by construction the problem with topology changes. Earlier attempts to use 
a modified gauge action to fix topology did not lead to satisfactory results
since it was not possible to fix topology completely when values of the 
lattice spacings, say, $a\approx 0.1$fm were aimed at \cite{Fukaya:2005cw,Bietenholz:2005rd}.

As an alternative approach, the usage of a determinant ratio, 
\begin{equation}
R= \mathrm{det}\left[D_\mathrm{W}^2(-m_0)\right]
   /\mathrm{det}\left[D_\mathrm{W}^2(-m_0)+\mu^2\right]
\label{detratio}
\end{equation}
has been proposed in ref.~\cite{Fukaya:2006vs}. 
This constitutes another local modification of the gauge action 
since the masses used in eq.~(\ref{detratio}) are taken to be large. In particular, 
the bare quark mass of the Wilson Dirac operator $D_\mathrm{W}$ 
is taken to be negative which 
suppresses the occurrence of small eigenvalues, forbidding therefore topology 
changes. 
A number of overlap fermion simulations employing the determinant ratio
of eq.~(\ref{detratio}) have already been performed 
\cite{Aoki:2007pw,Aoki:2008ss,Aoki:2008tq,Ohki:2008ff}. 
An account of present simulations employing the 
determinant ratio is given in ref.~\cite{Hashimoto2008}. 

In this still rather new approach to lattice QCD a number of issues have 
to be 
clarified such as a test of the topological finite size effects
\cite{Brower:2003yx,Aoki:2007ka}, the 
ergodicity of the simulations and possibly long auto correlations. 
Nevertheless, I find this a very interesting way of obtaining the continuum 
limit: in the continuum, the total topological charge will average out to zero, 
while local topological charges will, of course, still appear. 
Thus, it is a valid and intriguing approach to fix topology to zero 
from the very beginning and see how the system behaves towards the 
continuum limit. From my point of view, this offers a nice alternative
for QCD simulations.

\vspace*{0.5cm}
\subsection{Other approaches}
\vspace*{0.5cm}

There are more alternatives of lattice QCD formulations, 
such as FLIC fermions \cite{Boinpolli:2007zz}, 
chirally improved fermions \cite{Gattringer:2008td},
perfect action fermions \cite{Hasenfratz:2006xi} and Hyp-link  
smearing techniques \cite{Schaefer:2007dc}. 
Simulations with these kind of fermions have not yet reached as 
ambitious parameter values as many of the large collaborations employ 
and which use the fermion formulations discussed above.

\subsection{Summary of action discussion}

There are a number of interesting fermion actions on the market. 
Each of them has certain shortcomings the most important of which are: \\ 
{\em $\mathrm{O}(a)$-improved Wilson fermions}: breaking of chiral symmetry, 
non-perturbative operator improvement;\\
{\em rooted staggered fermions}: taste breaking, non-local lattice action;\\
{\em twisted mass fermions}: breaking of chiral symmetry, isospin breaking;\\
{\em overlap fermion}: expense of simulation;\\
{\em domain wall fermions}: expense of simulation and breaking of chirality;\\
{\em smearing}: effects of high levels of smearing;\\
{\em fixed topology}: topological finite size effects. 

It seems that there is no ideal action 
which is obvious to select. Therefore, just to re-iterate, 
a universality test showing which 
of these actions lead to consistent continuum limit values 
is a necessity. 

\section{Chiral perturbation theory}

The fact that nowadays pseudo scalar masses below 300MeV can be reached, 
offers the possibility to confront the numerically obtained
data with the corresponding expressions from chiral perturbation 
theory. It is important to realize that the values 
for the low energy constants obtained from fits to chiral 
perturbation theory can be used in return for many phenomenological 
applications by inserting them into the relevant formulae of 
chiral perturbation theory. Thus a reliable and precise calculation 
of the low energy constants is a most valuable outcome of lattice simulations. 
In consequence, studying the mass dependence of many physical quantities
in lattice QCD is important and, of course, actively pursued. 

When discussing chiral perturbation theory in the context of lattice 
simulations one has to specify the setup in which the discussion is 
taking place. There are essentially three cases, (i) SU(2) chiral perturbation 
theory applicable to $N_f=2$ mass degenerate quarks, (ii) the corresponding 
SU(3) case and (iii) the case where we have light, mass-degenerate up and 
down quarks and a strange quark at its physical value. 

\subsection{SU(2) chiral perturbation theory} 

The classical quantities to confront with chiral perturbation theory are the 
pseudo scalar mass and the pseudo scalar decay constant which can be
determined very precisely from lattice QCD simulations.
When a range of quark masses is considered that corresponds to 
an interval of pseudo scalar masses of $250 \lesssim m_\mathrm{PS} \lesssim 450$MeV 
then it seems that the 1-loop chiral perturbation theory formula 
(see refs.~\cite{Gasser:1986vb,Colangelo:2005gd} for the adequate fitting formulae)
is applicable as seen in the examples shown in 
fig.~\ref{fig:lec1} (from ETMC) and fig.~\ref{fig:lec2} (from the JLQCD collaboration). 
In fact, the data are 
described by the 1-loop expression so well that there is no room for 
any sensitivity for the 2-loop corrections. 
Fig.~\ref{fig:lec2} also demonstrates that going beyond 
pseudo scalar masses of, say, $450$MeV the chiral fits become problematic
since fits using alternative expansion parameters
lead to significant differences. 

A conclusion that for SU(2) chiral perturbation theory the 1-loop 
formula for the above given mass range is satisfactory is, however, 
possibly pre-mature. Examples are the  vector and the charged radii 
of the pseudo scalar particle as computed by 
JLQCD \cite{Kaneko:2007nf} and ETMC \cite{Simula:2007fa}. 
Here, a 1-loop chiral perturbation theory
formula cannot describe the data appropriately and a NNLO correction has 
to be taken into account.
This holds true, even if the same range of pseudo scalar masses 
is used for which the quark mass dependence of 
$f_\mathrm{PS}$ and $m_\mathrm{PS}$ are described perfectly by NLO 
chiral perturbation theory. 

It is an open question, as to whether the failure to describe the pion radii
by the 1-loop expression of chiral perturbation is due to the fact 
that even for $250 \lesssim m_\mathrm{PS} \lesssim 450$MeV the 
2-loop correction is necessary or, 
whether the zero quark mass asymptotics of different observables is
qualitatively different. 
To answer this question, presumably many quantities have to be 
fitted simultaneously such that the 2-loop low energy constants 
can be reliably determined. 
Having the LECs in our hand, it will then become possible to 
quantify the 2-loop corrections for given values of the pseudo scalar 
mass.

\subsection{SU(3) chiral perturbation theory}

Up to my knowledge there has been so far no attempt to perform 
dedicated simulations with $N_f=3$ mass degenerate quarks to compare 
with chiral perturbation theory \cite{Gasser:1982ap,Gasser:1983yg}. 
In my opinion such simulations would, 
however, be important for two reasons. The first is obviously that we want 
to compare the low energy constants from a SU(3) chiral perturbation theory
fit to the corresponding 
case of SU(2). The second is that for the non-perturbative renormalization 
of $N_f=2+1$ lattice QCD simulations
preferably a massless renormalization scheme should be used which requires
simulations at a number of quark masses employing 
$N_f=3$ mass degenerate flavours and then to perform an extrapolation to the chiral point. 
Such simulations would automatically generate the set of data to confront to
SU(3) chiral perturbation theory and are planned by e.g. 
by the MILC collaboration \cite{Hellerpriv}.

\subsection{$N_f=2+1$}

In most simulations we have the situation that 2 mass degenerate up and down 
quarks and a strange quark close to its physical value are employed. The simulations 
are then performed by varying the light quark masses while keeping 
the strange quark mass roughly constant in physical units. 

\begin{figure}[t]
  \centering
  \subfigure[Applying NLO SU(3) chiral perturbation theory   
   to the Kaon decay constant $f_K$ (from the PACS-CS 
   collaboration). A clear discrepancy between the numerical 
   data and chiral perturbation theory predictions can be observed.    
  \label{fig:chiral1}]
  {\includegraphics[width=0.45\linewidth]{fk_su3.eps}}\quad\quad
  \subfigure[Application of continuum NNLO chiral
    perturbation theory to staggered fermion simulations at 
    a value of the lattice spacing of $a\approx 0.06$fm.
   \label{fig:chiral2}]
  {\includegraphics[width=0.38\linewidth]{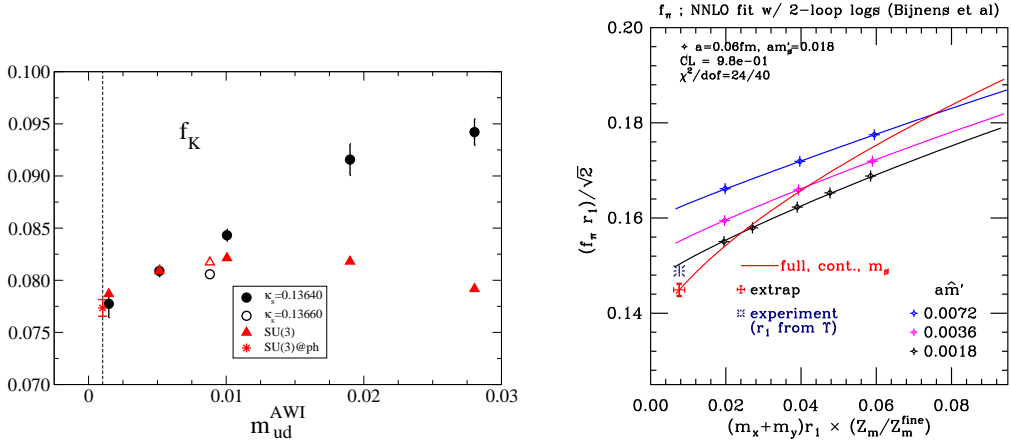}}
  \caption{Chiral perturbation theory for $N_f=2+1$.}
  \label{fig:chiral}
\end{figure}

Attempts to describe then the mass dependence of the pseudo scalar decay 
constant up to the Kaon scale by SU(3) chiral perturbation theory \cite{Gasser:1984gg}
are not successful. In 
fig.~\ref{fig:chiral1} we give an example from the PACS-CS collaboration 
\cite{Aoki:2008sm,Ukita:2008mq,Kadoh:2008sq} which shows the comparison of the 
Kaon decay constant $f_K$ to NLO chiral perturbation theory. 
Clearly, there is a large discrepancy 
between the measured values from the lattice simulations and 
chiral perturbation theory.  
The description of the numerical data breaks down rather early in the 
quark mass and any attempt to extend the formulae up to the strange 
quark mass fails. Such a behaviour is also observed by other collaborations: 
the RBC-UKQCD collaboration \cite{Allton:2008pn,Scholz:2008uv} uses 
an effective Kaon chiral perturbation theory to fix the problem;
in the case of staggered fermions, the lattice artefact corrections are taken 
into account \cite{Aubin:2005aq} which enlarges, however, the set of  
parameters to be fitted substantially. But still, a NLO formula from chiral perturbation 
theory does not seem to be sufficient to describe the numerical data 
up to the Kaon scale.
It is therefore tried to use 
\cite{Bernardchiralpriv}
{\em continuum} NNLO chiral perturbation theory for the smallest value
of the lattice spacing of $a=0.06$fm for staggered fermion simulations. 
This is shown in fig.~\ref{fig:chiral2}. 
Since further simulations at an even smaller value 
of the lattice spacing are planned (or even already ongoing) this offers a 
nice way to reduce the number of free parameters and test the applicability 
of chiral perturbation theory in the continuum. 

To summarize, for $N_f=2$ mass degenerate flavours of quarks 
chiral perturbation theory seems to work very well, although it is 
not clear whether the NLO formula is applicable for all quantities. 
The situation 
when adding the strange quark mass is problematic and a simple application 
of chiral perturbation theory does not work. Here, some input 
and interaction with experts from chiral perturbation theory is 
highly welcome.

\section{Some additional issues}

\subsection{Mixed actions}

In order to compute physical observables, often a mixed action approach 
is used. Here, the kind of lattice fermions used for generating the 
configurations, the {\em sea quarks}, is different from the kind of 
lattice fermions used to compute the propagators, {\em the valence quarks}. 
Such a procedure is particularly useful, if we think of 
computationally very expensive fermions such as overlap or 
domain wall fermions
and if 'wrong chirality' mixings in the twisted mass
regularization are to be tackled \cite{Frezzotti:2003xj}. 
In order to relate valence and sea quarks, an appropriate 
matching condition ought to be applied. To this end, typically the bare 
parameters of the valence quark action is tuned in such a way that 
the pseudo scalar mass of sea and valence quarks match. Keeping such a 
matching condition towards the continuum limit will then give a unitary 
theory in the continuum limit, at least if the lattice sea and valence  
quark actions were unitary by themselves. 

Although such a mixed action approach is therefore conceptually sound, 
it is not studied in great detail yet. In particular, for any 
non-zero value of the lattice spacing very special lattice artefacts 
can appear. For example, the scalar correlator can become negative 
and the 
lattice spacing corrections towards the continuum limit can get 
additional contributions from the fact that the sea and the valence 
quark masses are different 
\cite{Bar:2002nr,Bar:2005tu,Bar:2003mh,Prelovsek:2004jp,Golterman:2005xa,Bar:2006zj,Aubin:2008wk}. 

To illustrate that care has to be taken in this mixed action approach
I give two examples. The first is a calculation of overlap valence 
quarks on a maximally twisted mass sea \cite{Garron:2007tj} at a 
value of the lattice spacing of about $a\approx 0.09$fm. 
While matching the pseudo scalar
mass, the values of the pseudo scalar decay constants show a remarkable 
discrepancy at the matching point, $af_\mathrm{PS}^\mathrm{tm}=0.0646(4)$ while
$af_\mathrm{PS}^\mathrm{overlap}=0.077(4)$. 
Another example is a domain wall valence computation on a rooted 
staggered sea \cite{WalkerLoud:2008bp} at a value of the lattice spacing 
of about $a\approx 0.124$fm. 
Again matching the pseudo scalar mass, a significant difference in the
nucleon mass is found: $aM_\mathrm{nucleon}^\mathrm{staggered}=0.723(6)$ 
while $aM_\mathrm{nucleon}^\mathrm{domain wall}=0.696(7)$. 
Since in the continuum limit the values of 
physical observables have to agree, these two examples hint at rather 
large lattice artefacts appearing in a mixed action setup. Thus, a careful 
check of lattice artefacts will be very useful and is almost mandatory. 
Note, however, that for closely related actions such as Osterwalder-Seiler
quarks \cite{Osterwalder:1977pc} 
on a twisted mass sea \cite{Vladikas2008} or unrooted staggered valence 
on rooted staggered sea fermions, 
physical observables seem to match better. 

Fortunately, the lattice spacing effects in a number of mixed action 
formulations have been analyzed in lattice chiral perturbation theory
\cite{Bar:2002nr,Bar:2005tu,Bar:2003mh,Golterman:2005xa,Chen:2005ab,Chen:2006wf,Chen:2007ug}. 
These formulae can and have been used to describe the numerical data. 

\subsection{Non-perturbative renormalization} 

Doubtlessly, non-perturbative renormalization is a necessity 
in lattice QCD simulations. This can be illustrated with the example
of the strange quark mass,  which obtains a value of 
$m_\mathrm{strange}^\mathrm{perturbative}=72\pm 2\pm 9$MeV while
$m_\mathrm{strange}^\mathrm{nonperturbative}=105\pm 3\pm 9$MeV 
\cite{Blossier:2007vv}. 
Note that the values of the perturbative renormalized strange quark mass
taken here from ETMC is fully consistent with the 
corresponding PACS-CS result \cite{Aoki:2008sm}. 
A similar picture emerges for the 
light quark masses. 

In order to obtain the non-perturbatively evaluated renormalization 
constants in a {\em mass independent} renormalization scheme, 
either the RI-MOM \cite{Martinelli:1994ty} 
or the Schr\"odinger functional (SF) 
scheme \cite{Luscher:1992an,Jansen:1995ck}
can be used. In the former case, an extrapolation to the chiral 
limit has to be performed, while in the second case the theory can be 
simulated  
directly at or close to zero quark masses. 

For the case of $N_f=2$ mass degenerate quarks such procedures have
been already successfully applied, see \cite{Sommer:2006sj}. 
For the case of $N_f=2+1$, there is the additional complication that 
the strange quark mass is kept close to its physical value. Therefore, 
in order to obtain a massless renormalization scheme, additional 
runs with $N_f=3$ mass degenerate quarks would have to be performed in principle. 
Such simulations are not available yet but e.g. 
MILC is planning such runs \cite{Hellerpriv}. 
As mentioned above such simulations have the additional advantage 
that SU(3) chiral perturbation theory can be checked and eventually the 
SU(3) low energy constants be extracted. 

For the time being, collaborations such as RBC-UKQCD try to estimate 
the systematic effects coming from a fixed and rather large strange 
quark mass and add this as a systematic error in the renormalization 
constants \cite{Allton:2008pn}. However, this needs an explicit check. 
Also, first investigations with the SF scheme and $N_f=3$ flavours of 
quarks are under way \cite{Taniguchi2008}.
For theoretical discussion of SF boundary conditions at this conference
see \cite{Sint2008,Lopez:2008ns}.

\subsection{Effects of strange quark}

Often a question is asked whether the results from 
$N_f=2$ flavours of quarks are reliable since the strange 
quark is neglected and taken only as a valence quark in the calculation 
of various observables. 

\begin{figure}[t]
  \subfigure[Omega mass at fixed value $r_0m_\mathrm{PS}$ as a function 
   of $a^2$. Data are from MILC (having a dynamical strange) and ETMC 
   (with only up and down sea quarks).
  \label{fig:omega}]
  {\includegraphics[width=0.55\linewidth]{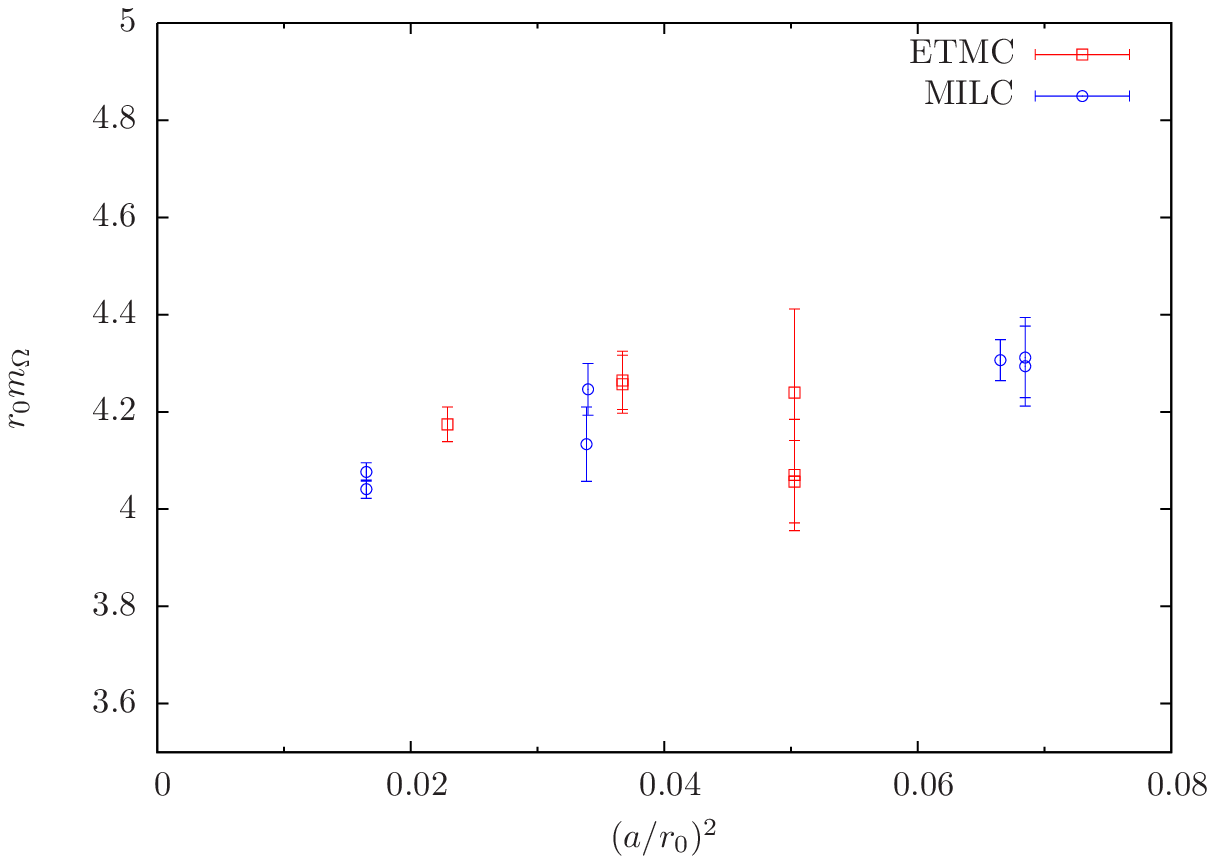}}\quad
  \subfigure[Finite size effects for $g_A$ as determined by the 
    QCDSF collaboration. The ratio of the infinite volume 
    value $g_A$ and its finite volume analogue $g_A(L)$ can reveal
    large finite size effects at the 15\%-20\% level.  
   \label{fig:fse}]
  {\includegraphics[width=0.37\linewidth,angle=90,angle=90,angle=90]{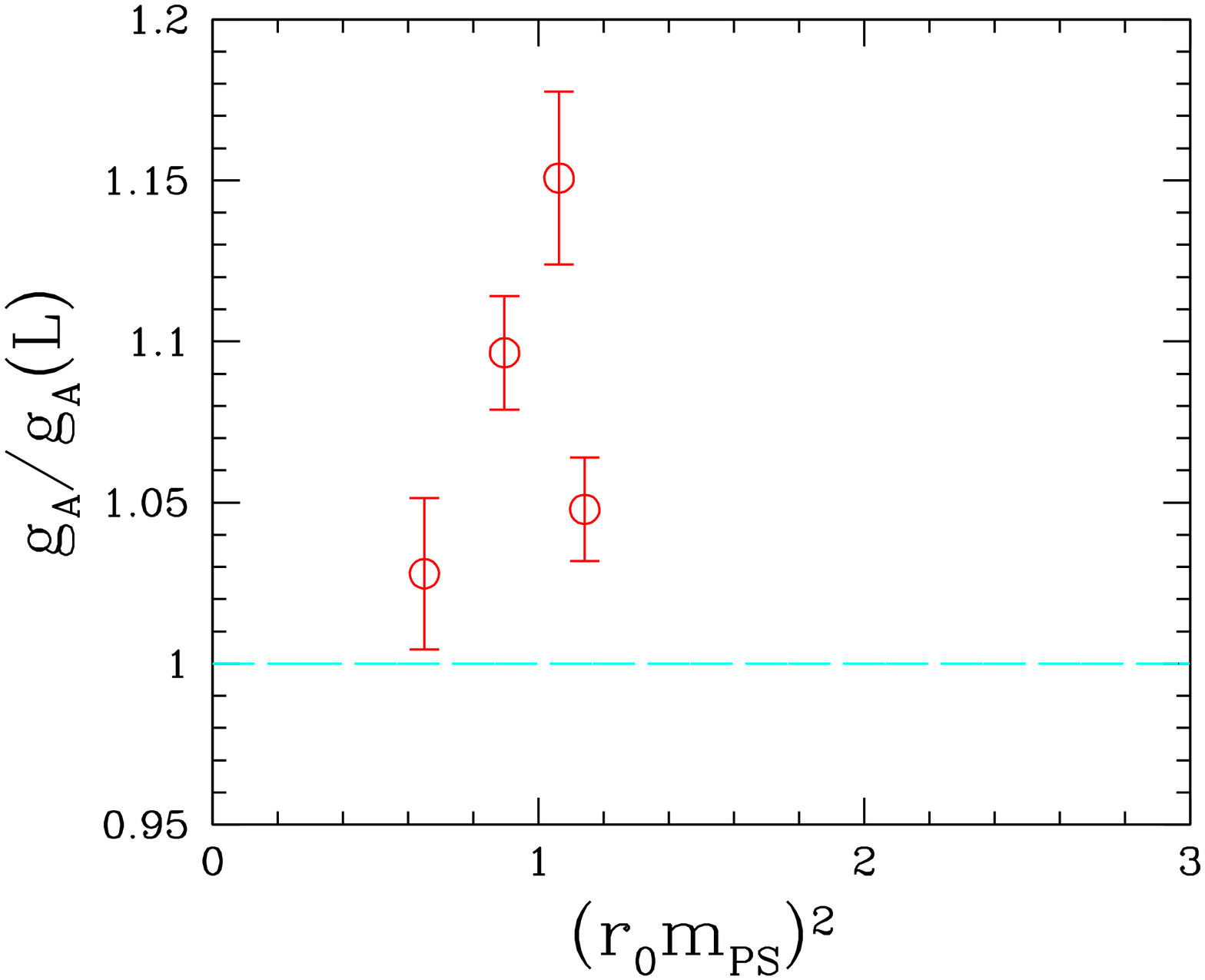}}
  \caption{Omega mass and finite size effects of $g_A$.}
  \label{fig:omegafse}
\end{figure}

In order to see any effects of a dynamical strange quark, 
a most sensitive quantity should be 
the $\Omega$ baryon which consists of three strange quarks and has 
no strong decay.
In fig.~\ref{fig:omega} results from computations of MILC \cite{Bernard:2007ux} 
(which has a dynamical strange quark) 
and ETMC \cite{Drach2008} (which uses only up and down quarks in the sea) 
for the 
$\Omega$ baryon are compared at various values of the lattice spacing keeping  
$r_0m_\mathrm{PS}$ fixed. Within the error bars, no
evidence of an effect of the strange quark mass is seen. To reveal such 
an effect, presumably the error bars would have to shrink substantially.
Up to my knowledge, 
also in other quantities no evidence (with possibly the exception 
of $f_{D_{s}}$) of the relevance of a dynamical 
strange quark has been observed so far and it will be interesting to 
see in the future whether and when such effects show up. 

\subsection{Finite size effects}

Simple mesonic quantities such as $m_\mathrm{PS}$ and $f_\mathrm{PS}$
are computed so precisely in present day numerical simulations that effects
of a finite volume are clearly visible and become a dominant systematic 
error. However, it seems that the analysis performed in ref.~\cite{Colangelo:2005gd}
provides an adequate description of the finite size effects 
for $m_\mathrm{PS}$ and $f_\mathrm{PS}$ as confirmed by many 
groups. In particular, if values of $m_\mathrm{PS}L\gtrsim 3.5$ are used, while
keeping $L$ itself large enough to avoid squeezing 
effects of the wave function \cite{Fukugita:1991hw,Fukugita:1992jj}, 
the finite size effects are at the percent level and can be fully controlled
by applying the formulae of ref.~\cite{Colangelo:2005gd}.

However, the nice results for these basic mesonic quantities
cannot be taken over automatically to other 
quantities. As demonstrated in fig.~\ref{fig:fse} by the example of $g_A$ (discussed by
the QCDSF collaboration \cite{Khan:2006de}), other quantities may have finite 
volume effects that can reach 15\%-20\%. Similar finite volume effects were 
observed for the ratio $g_A/g_V$ by the RBC-UKQCD collaboration.
Thus, finite volume effects need to be carefully investigated on a case
by case study.

\subsection{Topology}

\begin{figure}[t]
  \subfigure[Topological susceptibility from the RBC-UKQCD collaboration 
   in the approach to the chiral limit. In particular, the data on the larger
   lattice indicate that the topological susceptibility will vanish 
   when massless quarks are reached. 
  \label{fig:toposus}]
  {\includegraphics[width=0.45\linewidth,angle=90,angle=90,angle=90]{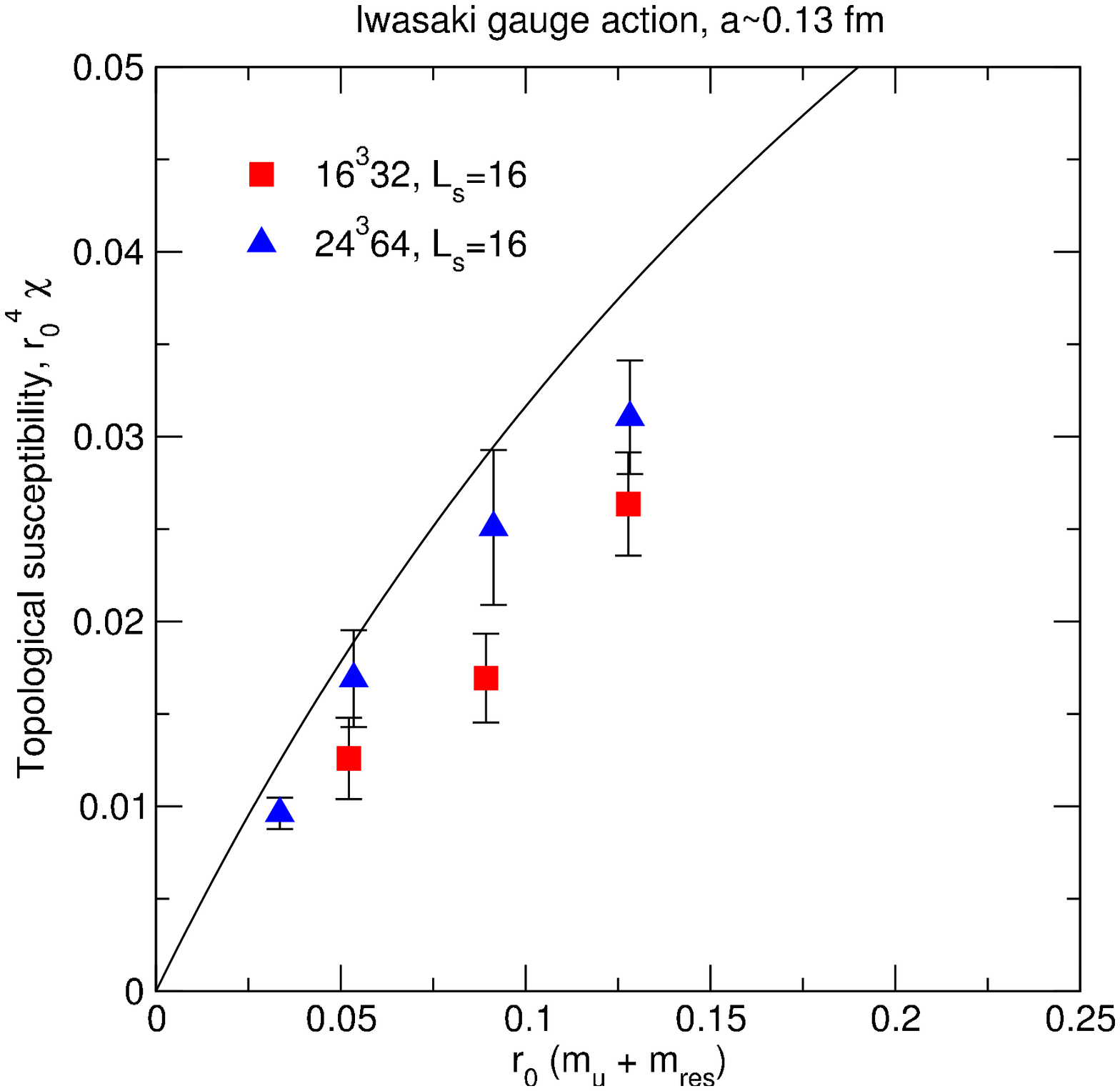}}\quad
  \subfigure[The $\eta_2$ mass (the analogue of the $\eta'$ mass 
    for two flavours of quarks) as a function of the pseudo scalar mass. 
    The flatness in the mass dependence allows an estimate at the 
    physical point of $\eta_2\approx 865$MeV.
   \label{fig:eta}]
  {\includegraphics[width=0.40\linewidth]{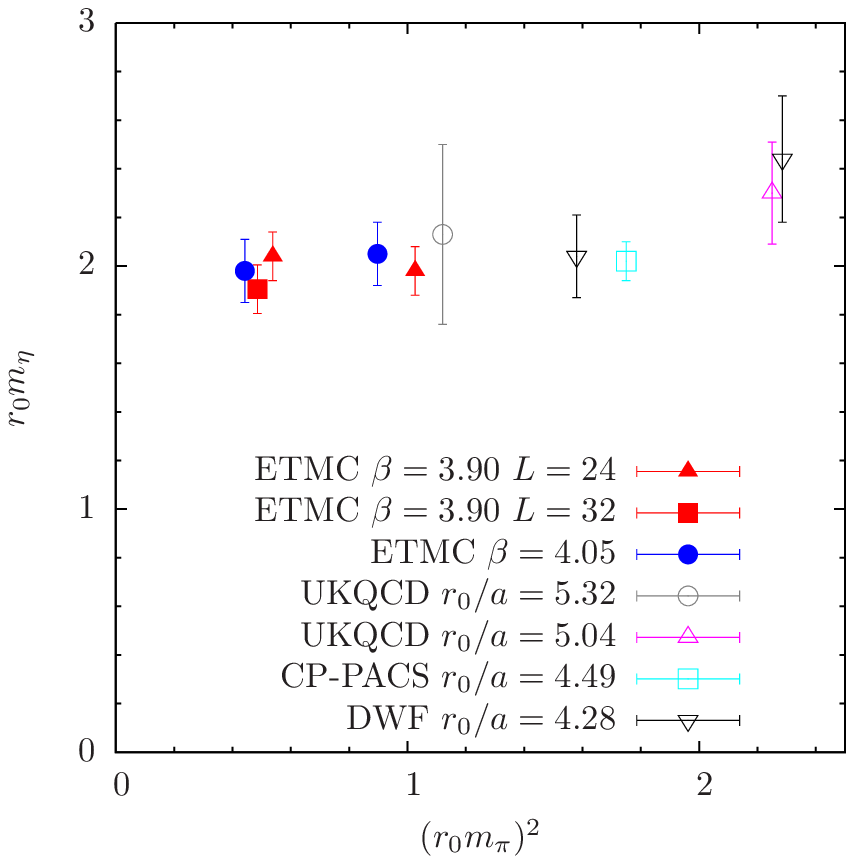}}
  \caption{The topological susceptibility and the $\eta_2$ mass.}
  \label{fig:topoeta}
\end{figure}

The question of topology on the lattice is one of the most interesting 
and difficult one to address. However, in the last years, we have seen 
much progress in this direction 
\cite{Chiu:2008kt,DeGrand:2007tx,Wenger2008,Bernard:2007ez,Durr:2006ky}. 
As only one example I show in fig.~\ref{fig:toposus} the mass dependence of the 
topological susceptibility towards the chiral limit as obtained by the 
RBC-UKQCD collaboration. 
The point I want to make here is that the topological susceptibility 
shows the right behaviour towards the chiral limit in that it  
vanishes as we approach massless quarks. This behaviour is also seen 
from other 
formulations, see the references given above. 

As a second example for a quantity which is directly related to topology, I 
show in fig.~\ref{fig:eta} the $\eta_2$ mass from $N_f=2$ 
simulations. The $\eta_2$ mass is the analogue of the $\eta'$ mass for $N_f=2+1$. 
Using the much improved algorithms for the simulations, advances  
of computing disconnected diagrams as well as new methods, it is possible
to reach small values of the pseudo scalar mass and small errors 
for this difficult to 
compute quantity. The graph, compiled by ETMC, ref.~\cite{Jansen:2008wv},  
reveals a basically flat behaviour of the $\eta_2$ mass as a function 
of the pseudo scalar mass and confirms that at the physical point a value
of the $\eta_2$ mass of $M_{\eta_2}\approx 865$MeV can be extracted thus 
showing a large contribution to the mass by topological effects. 
Note that with $N_f=1$ dynamical quarks \cite{Farchioni:2008na}
the corresponding $\eta^\prime$ ($\eta_1$) mass comes out to be 
330(20) MeV, in agreement with the Witten-Veneziano formula.

\subsection{Getting social}

As a last section in this discussion of a number of selected topics 
concerning lattice QCD simulations, I would like to address the communication 
within our lattice community. Although a strong competition between 
various ``dynasties'' of international collaborations is 
very welcome, there are, in my opinion, some easy to realize 
ways to homogenize our efforts and give therefore a more coherent picture 
to the outside world. 

\noindent {\em ILDG:} It would be very good if all collaborations were willing to 
upload their configurations to the ILDG. Although there an initial threshold 
effort, afterwards using ILDG tools become routine and a number of collaborations 
employ the ILDG tools successfully and efficiently already in their daily work. 
The usage of the ILDG format
for the configurations allows for an easy exchange of configurations that can 
provide valuable cross checks among different collaborations. 
See refs.~\cite{Ukawa:2004he,Jansen:2006ks,DeTar:2007au,Yoshie2008} for overviews 
of ILDG activities.

\noindent {\em Codes:} The algorithms used for present days simulations have become
rather complicated and it is no longer true that it takes a few days to write a 
Hybrid Monte Carlo code that includes state of the art improvements from scratch. 
In such a situation, it would be very good
if such complicated codes could be made available to the lattice community, preferably 
as an open source platform such that useful additions can be implemented. Examples of 
published codes are \cite{Luscherhp,Joo:2008bf,Borici:2006ch}.
Other collaborations are encouraged to follow up on these examples. 

\noindent {\em Details:} As discussed at length in the preceding sections, the 
results from lattice simulations suffer from a number of systematic effects 
that have to be controlled as well as possible. In order to be able to judge whether
this has been achieved in the work of a particular collaboration, it would therefore
be necessary to know about the details of the simulation, the analysis and 
the estimates of the systematic effects. Therefore, I would like to encourage 
all the collaborations to not only publish high gloss papers with final results, 
but also technical papers with all technical details of their work. This will 
allow everybody to judge and cross-check the results, but may also teach us 
about the techniques and 
whether they are of interest for other collaborations. 
In addition, it would be very useful to publish tables of raw data. Another aspect
is to perform blind analyses in order to avoid possible human interfaced biases.

\section{Conclusions}

The main message of this proceeding is the very substantial 
progress lattice field theory has achieved in the last years. Due to 
algorithmic breakthroughs, (see fig.~\ref{fig:cost}),
as the major factor in combination 
with a significant increase of 
super computer power and conceptual developments, several international
collaborations are nowadays performing simulations that were unthinkable 
even
a few years ago.  
In particular, in lattice QCD we are now reaching lattice spacing values 
of $a\approx 0.05$fm, pseudo scalar masses of about $250$MeV and below and 
box sizes with linear extent of $L\approx 3$fm.
Using $\mathrm{O}(a)$ improved lattice actions allows 
eventually for controlled
continuum, chiral and infinite volume extrapolations. 
Fig.~\ref{fig:parameters} summarizes the values of the lattice spacing and 
pseudo scalar masses that are covered in typical simulations presently. 

Examples of physical results that are available already now and which are 
computed as continuum quantities with systematic errors taken into 
account are the 
baryon spectrum as represented in fig.~\ref{fig:baryonspectrum} and 
the precise determinations of several low energy constants, see fig.~\ref{fig:lec}.
Many more physical results are to be expected in the near future since 
much of the raw data of lattice QCD, the dynamically generated gauge field 
configurations, exist already or will be generated soon. They are partly 
stored on the 
International Data Grid where they are often freely available. 

Despite this undeniable progress, caveats remain. The actions employed for the 
dynamical simulations lead to systematic errors that need to be controlled such 
as explicit breaking of chiral symmetry, isospin and taste breaking, 
high-level of smearing  and 
non-locality. In addition, the various actions show  
different kind of lattice artefacts. 
Therefore, a check is needed 
that different lattice fermion 
formulations lead to the same continuum limit values and a test of this 
universality is, in my opinion, one of the most urgent demands in lattice QCD. 
This problem is highly non-trivial as fig.~\ref{fig:fps1} demonstrates: here 
a compilation of many available lattice results for the pseudo scalar 
decay constant reveals a warning:  
no common scaling is observed when different 
lattice fermions are considered. This is in contrast to 
the nucleon mass of fig.~\ref{fig:nucleon1} where a good overall lattice
spacing scaling can be observed.

There are also a number of open questions that remain to be clarified: 
how to use chiral perturbation theory when a dynamical strange is 
fixed at its physical value? Related to this is the question 
of how best to extrapolate e.g. baryons and other quantities 
to the physical point. How about the non-perturbative
renormalization in the case of $N_f=2+1$ flavours? 
Can we control the finite volume effects 
for quantities different from simple mesonic observables? 
Should we include the charm as a dynamical degree of freedom and what 
will be the lattice artefacts?
How to best treat unstable particles in lattice QCD?
These are some of the challenges that the lattice QCD community has to 
address and solve.

Although there are for sure still a number of obstacles 
to overcome, lattice QCD simulations have finally become realistic. 
The physics 
coming out of such simulations have therefore to be discussed prudently. 
And, just to finish, it is maybe indeed the time now to make a serious 
effort towards a lattice particle data booklet. 

\section*{Acknowledgments}

This proceedings contribution would not have been possible 
without the help of many of my colleagues. In particular, I am most 
grateful to G. Herdoiza, I. Montvay, C. Urbach and M. Wagner
for helping me with generating plots and investing extra efforts to 
clarify physics questions such as the effect of stout smearing, 
the scaling behaviour of observables and the relation of the first order 
phase transition and the stability criterion. 

I am also indebted to the colleagues who sent data before publication 
or very useful comments to me. I would like to mention in 
particular C. Bernard, P. Boyle, S. Capitani, N. Christ, M. Creutz, 
C. DeTar, 
M. Golterman, S. Gottlieb, S. Hashimoto, C. Jung, Y. Kuramashi, D. Leinweber, 
D. Pleiter, G. Schierholz, E. Scholz, 
Y. Shamir, S. Sharpe, D. Toussaint and A. Ukawa
for their very prompt and helpful replies. 

Finally, I thank G. Herdoiza, I. Montvay, D. Renner, 
G. Rossi and C. Urbach for a careful
reading of and comments on this manuscript. 

%

\end{document}